\documentclass{cupconf}
\usepackage{epsfig}
\usepackage{natbib}

\newcommand{\apj}{\textit{ApJ}}
\newcommand{\apjs}{\textit{ApJS}}
\newcommand{\apjl}{\textit{ApJL}}

\newcommand{\aap}{\textit{A{\&}A}}
\newcommand{\aaps}{\textit{A{\&}AS}}
\newcommand{\mnras}{\textit{MNRAS}}
\newcommand{\aj}{\textit{AJ}}

\title[X-ray Tour of Massive Star-forming Regions]{An X-ray Tour of Massive Star-forming Regions with Chandra}

\author[L. K. Townsley]%
{L\ls E\ls I\ls S\ls A\ns K.\ns T\ls O\ls W\ls N\ls S\ls L\ls E\ls Y$^1$%
}

\affiliation{$^1$Department of Astronomy and Astrophysics, Pennsylvania State University, 525 Davey Laboratory, University Park, PA 16802, USA}

\pubyear{2006}
\volume{??}
\pagerange{??}
\date{?? and in revised form ??}
\begin{document}

\maketitle

\begin{abstract}
The Chandra X-ray Observatory is providing fascinating new views of
massive star-forming regions, revealing all stages in the life cycles of
massive stars and their effects on their surroundings.  I present a {\em
Chandra} tour of some of the most famous of these regions: M17, NGC~3576,
W3, Tr14 in Carina, and 30~Doradus.  {\em Chandra} highlights the physical
processes that characterize the lives of these clusters, from the ionizing
sources of ultracompact HII regions (W3) to superbubbles so large that
they shape our views of galaxies (30~Dor).  X-ray observations usually
reveal hundreds of pre-main sequence (lower-mass) stars accompanying
the OB stars that power these great HII region complexes, although in
one case (W3 North) this population is mysteriously absent.  The most
massive stars themselves are often anomalously hard X-ray emitters;
this may be a new indicator of close binarity.  These complexes are
sometimes suffused by soft diffuse X-rays (M17, NGC~3576), signatures
of multi-million-degree plasmas created by fast O-star winds.  In older
regions we see the X-ray remains of the deaths of massive stars that
stayed close to their birthplaces (Tr14, 30~Dor), exploding as cavity
supernovae within the superbubbles that these clusters created.
\end{abstract}

\firstsection 
\section{Revealing the Life Cycle of a Massive Stellar Cluster}

High-resolution X-ray images from the Chandra X-ray Observatory and
XMM-Newton elucidate all stages in the life cycles of massive stars
-- from ultracompact HII (UCHII) regions to supernova remnants --
and the effects that those massive stars have on their surroundings.
X-ray studies of massive star-forming regions (MSFRs) thus give insight
into the massive stars themselves, the accompanying lower-mass cluster
population, new generations of stars that may be triggered by the massive
cluster, interactions of massive star winds with themselves and with
the surrounding neutral medium, and the fate of massive stars that die
as cavity supernovae inside the wind-blown bubbles they created.

In the era of {\em Spitzer} and excellent ground-based near-infrared
(IR) data, why are X-ray studies of MSFRs important?  {\em Chandra}
routinely penetrates $A_V > 100$~mag of extinction with little confusion
or contamination, revealing young stellar populations in a manner that is
unbiased by the presence of inner disks around these stars.  These vast
pre-main sequence (pre-MS) populations are easily seen by {\em Chandra}
and the $17^\prime \times 17^\prime$ imaging array of its Advanced CCD
Imaging Spectrometer (ACIS-I); a typical ACIS-I observation of a nearby
($D<3$~kpc) MSFR, lasting 40--100 ksec, finds 500--1500 young stars,
tracing the initial mass function (IMF) down to $\sim 1$M$_\odot$.
These observations can increase the number of known cluster members
by as much as a factor of 50 in poorly-studied regions \citep[e.g.\
NGC~6357,][]{Wang06}.

In the first 7 months of 2006, at least 18 refereed papers on
X-ray observations of MSFRs (using {\em Chandra} or {\em XMM})
were published or submitted.  These include studies of 30~Doradus
\citep{Townsley06a,Townsley06b}, NGC~6334 \citep{Ezoe06}, $\sigma$
Orionis \citep{Franciosini06}, Westerlund~1 \citep{Skinner06,Muno06},
RCW~38 \citep{Wolk06}, Cepheus~B/OB3b \citep{Getman06}, NGC~6357
\citep{Wang06}, M17 \citep{Broos06}, NGC~2362 \citep{Delgado06},
NGC~2264 \citep{Flaccomio06,Rebull06}, Trumpler 16 in Carina
\citep{Sanchawala06}, NGC~6231 \citep{Sana06}, the Orion Nebula Cluster
\citep{Stassun06}, the Arches and Quintuplet clusters \citep{DWang06},
and W49A \citep{Tsujimoto06}.  Early {\em Chandra} and {\em XMM} studies
were summarized in \citet{Townsley03}; more recent work is described in
\citet{Feigelson06}.  Other MSFRs currently under study at Penn State
include RCW~49 \citep{Townsley04}, NGC~7538 \citep{Tsujimoto05}, W51A
\citep{Townsley04,TownsleyIAUS227}, and the Rosette Nebula (Wang et al.\
2006, in preparation).

This contribution begins with recent X-ray results on the young MSFRs
M17 and NGC~3576, two regions at different evolutionary stages but with
notably similar outflows of hot, X-ray-emitting gas.  Next I introduce
a new ACIS mosaic of the Westerhout 3 complex, showing how triggered
star formation in the same molecular cloud can result in very different
stellar clusters.  Then I address the rich and complicated point source
and diffuse X-ray emission seen towards Trumpler~14 in Carina.  Finally I
provide a first look at a new {\em Chandra} observation of 30 Doradus
obtained in early 2006.

\section{M17, The Omega Nebula:  An X-ray Champagne Flow}   

The Omega Nebula (M17) is the second-brightest HII region in the sky
(after the Orion Nebula).  Its OB cluster NGC~6618 probably has well over
5000 members, extending up to masses $M \sim 70+70 M_\odot$ (spectral
type O4V+O4V) in its central binary.  It is situated at the edge of one
of the Galaxy's most massive and dense molecular cloud cores, M17SW,
at a distance of 1.6~kpc \citep{Nielbock01}.  It is a blister HII region
viewed nearly edge-on, on the periphery of a giant molecular cloud (GMC)
containing an UCHII region with UV-excited water masers and active star
formation.  The ionization front of the HII region encounters the GMC
along two photodissociation regions, called the northern and southern
bars.  With an age of $<1$~Myr \citep{Hanson97} and no evolved stars,
M17 is one of the few bright MSFRs that has sufficient stellar wind
power to produce an X-ray outflow and yet is unlikely to have hosted
any supernovae.

\begin{figure} \centering

\psfig{figure=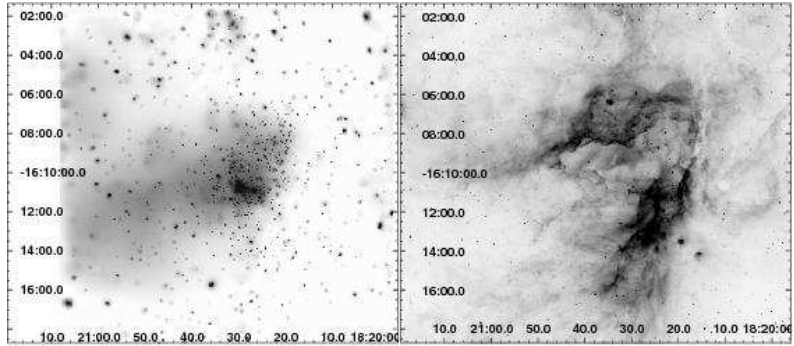,height=1.8in}
\small
\caption{\protect \small {\bf Left:}  Smoothed 0.5--2 keV ACIS-I image of
M17 ($\sim 8$~pc across), highlighting the soft diffuse X-ray emission.
Here and throughout this paper, coordinates are J2000 RA and Dec.
{\bf Right:} {\em Spitzer}/IRAC 5.8$\mu$m image of the same M17 field,
part of the GLIMPSE survey and provided by the GLIMPSE team.  
This pair of images highlights both the rich stellar population revealed
through X-ray/IR comparison studies and the complementarity of the two
Great Observatories for mapping diffuse emission in the region.
}
\normalsize
\label{fig:m17_xray_ir}
\end{figure}

We observed M17's central cluster NGC~6618 with {\em Chandra}/ACIS-I
for 40~ksec in March 2002.  The extensive soft diffuse emission seen
in M17 is rare; it was described by \citet{Townsley03} and is shown
in Figure~\ref{fig:m17_xray_ir}.  The 877 point sources detected in
this observation are described, catalogued, and compared to IR data by
\citet{Broos06}.  Fewer than 10\% of these point sources (in M17 and
in the other Galactic MSFRs described here) are unrelated to the MSFR
\citep[foreground stars and background AGN,][]{Wang06}.  {\em Chandra}
has a point spread function (PSF) that is sub-arcsecond on-axis, but
degrades radially to become many arcseconds wide at the field edge.
This effect is apparent in Figure~\ref{fig:m17_xray_ir} and in other
ACIS images shown in this paper.

Figure~\ref{fig:m17_xray_ir} (right) shows the 5.8$\mu$m GLIMPSE
image of M17 from the Spitzer Space Telescope's Infrared Array
Camera (IRAC).  The early {\em Chandra} and {\em Spitzer} images in
Figure~\ref{fig:m17_xray_ir} show dramatic and complex interacting
components of hot and cold material.  On the western side of the cluster,
the parsec-scale thermalized O-star winds shock the parental GMC,
triggering new massive star formation to the southwest.  To the east,
the winds escape the confinement of the blister HII region producing the
brightest known X-ray champagne flow, heating and chemically enriching
the ISM \citep{Townsley03}.

\begin{figure} \centering

\mbox{\psfig{figure=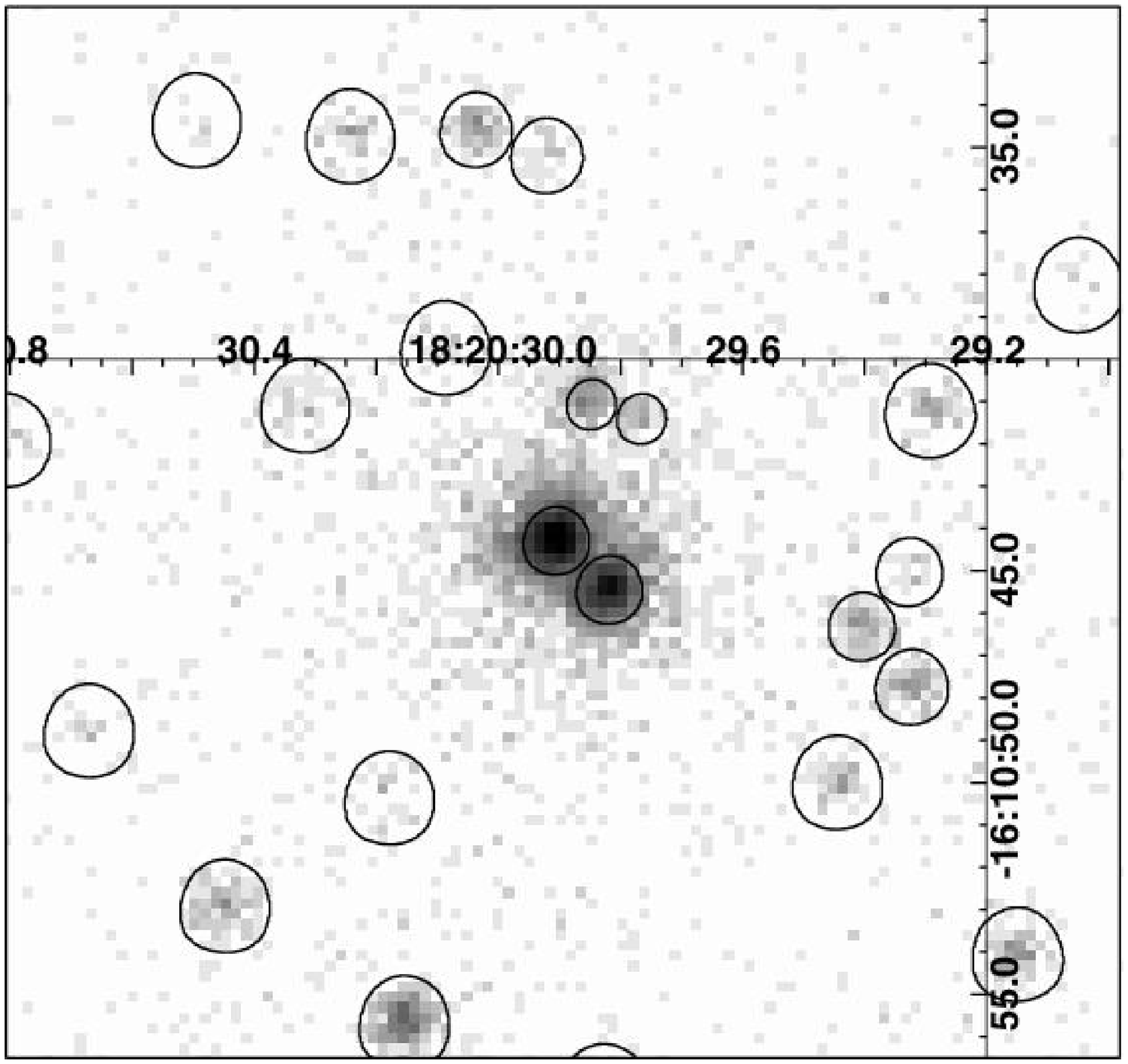,height=1.5in}
      \psfig{figure=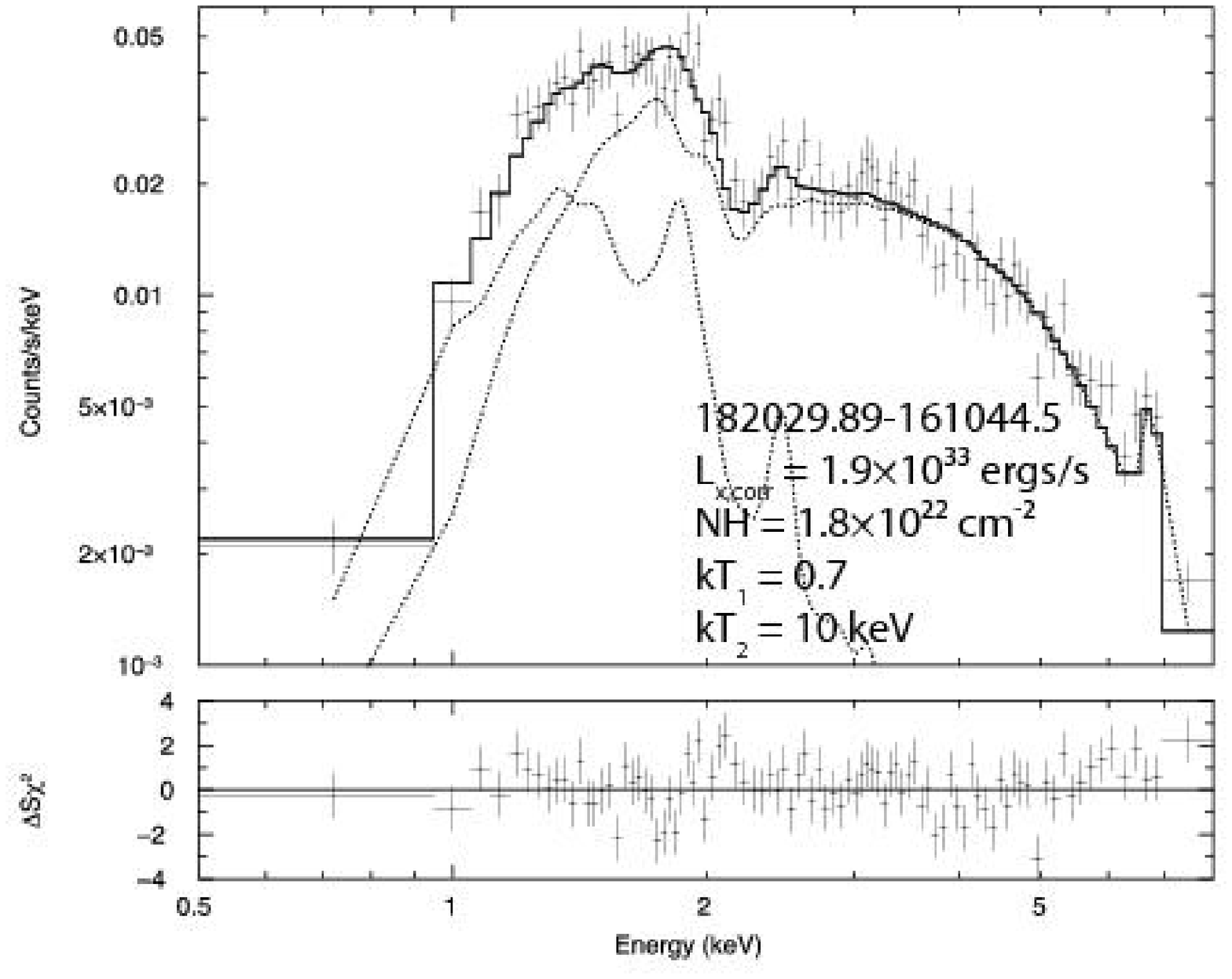,height=1.5in}
\hspace*{-0.2in}
      \psfig{figure=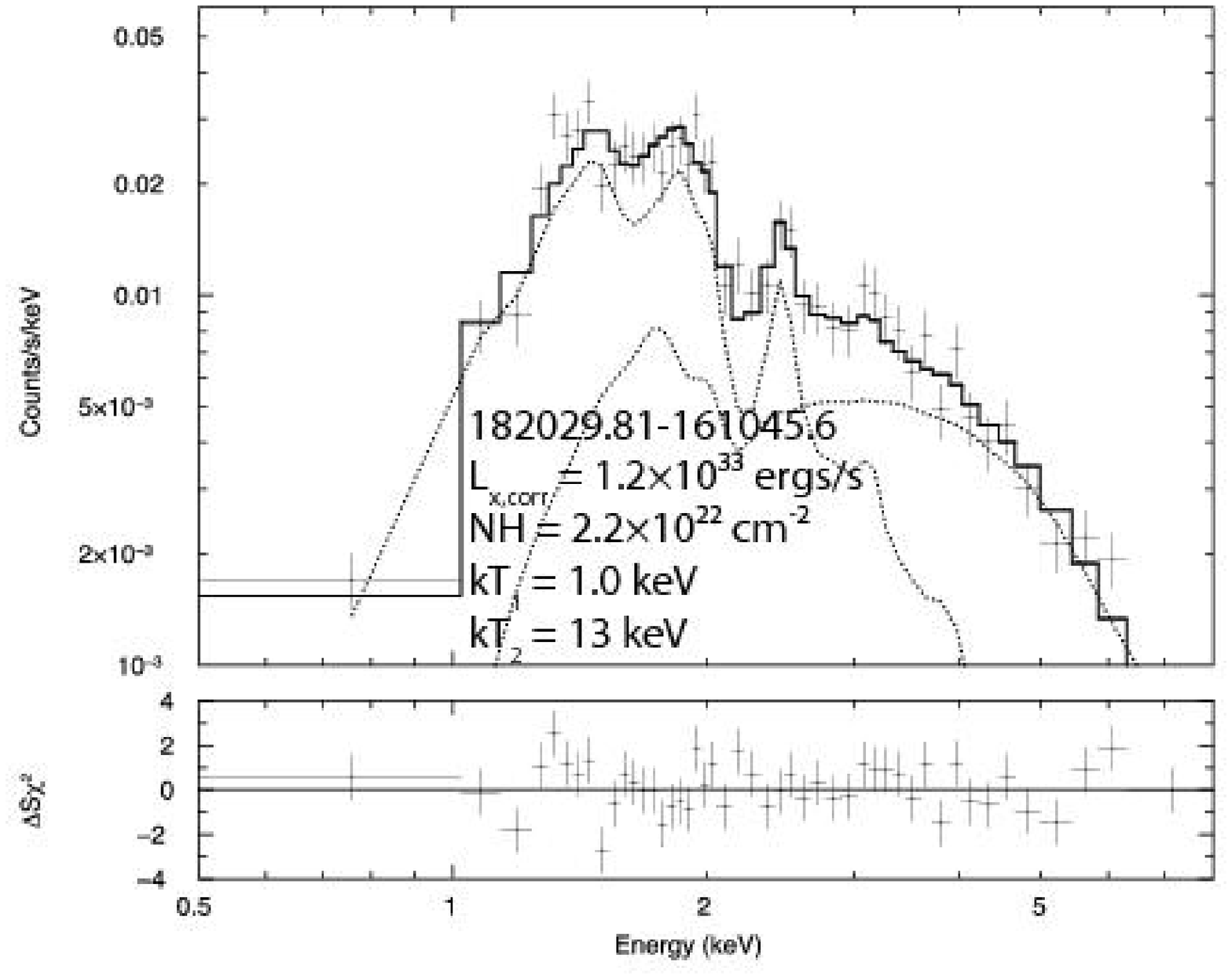,height=1.5in}
      }
\small
\caption{\protect \small {\bf Left:}  High-resolution ACIS image of the
central region of NGC~6618 in M17, highlighting the bright O4+O4 binary.
The nearly-circular polygons mark the photometric and spectral extraction
regions for each source.
{\bf Center and Right:}  ACIS spectra of the two O4 stars.  For each
source, the top panel shows the event data as a histogram, each component
of the model (convolved with the instrumental response) as dotted lines,
and the composite model as a solid line.  The lower panel shows the
fit residuals.
}
\normalsize
\label{fig:m17_O4_stars}
\end{figure}

{\em Chandra's} strongly-varying PSF and the ACIS-I camera's large
field of view also make it difficult to convey the sub-arcsecond imaging
quality in full-field smoothed images.  Figure~\ref{fig:m17_O4_stars}
(left) shows the central $\sim 25^{\prime\prime}$ of the M17 ACIS
image, binned to $0.25^{\prime\prime}$ pixels.  NGC~6618's central
binary is separated by $\sim 2^{\prime\prime}$ and is clearly resolved.
These are the brightest X-ray sources in the field \citep{Broos06}.
Figure~\ref{fig:m17_O4_stars} (right) shows their spectra and model fits.
Each source was fit using {\em XSPEC} \citep{Arnaud96} with a two-temperature {\it apec}
thermal plasma model \citep{Smith01}, including absorption.  The soft plasma components
in these O4 stars are of similar strength and are typical of O stars;
they are thought to be due to microshocks in the powerful stellar winds.
Both of these sources, however, show very hard thermal plasmas as well;
the brighter O4 star's spectrum is dominated by this hard component.
A reasonable explanation for this hard emission is that it comes
from colliding winds in an as-yet-unrecognized close binary system.
Since both O4 stars exhibit this hard emission, it is possible that the
O4+O4 binary actually consists of no fewer than 4 massive components.


\section{The Giant HII Region NGC~3576:  An M17 Analog?}

NGC~3576 is a Galactic giant HII region, located at a distance of 2.8~kpc
\citep{dePree99}.  It contains several known O and early B stars but
these are not sufficient to account for the Ly$\alpha$ ionizing photons
inferred from radio data \citep{Figueredo02}.  An embedded massive IR star
cluster is located at the edge of its GMC \citep{Persi94}.  It contains at
least 51 stars earlier than A0, most with large IR excesses.  It likely
includes as yet unrecognized O stars that would account for the number
of Ly$\alpha$ photons.

\begin{figure} \centering

\mbox{\psfig{figure=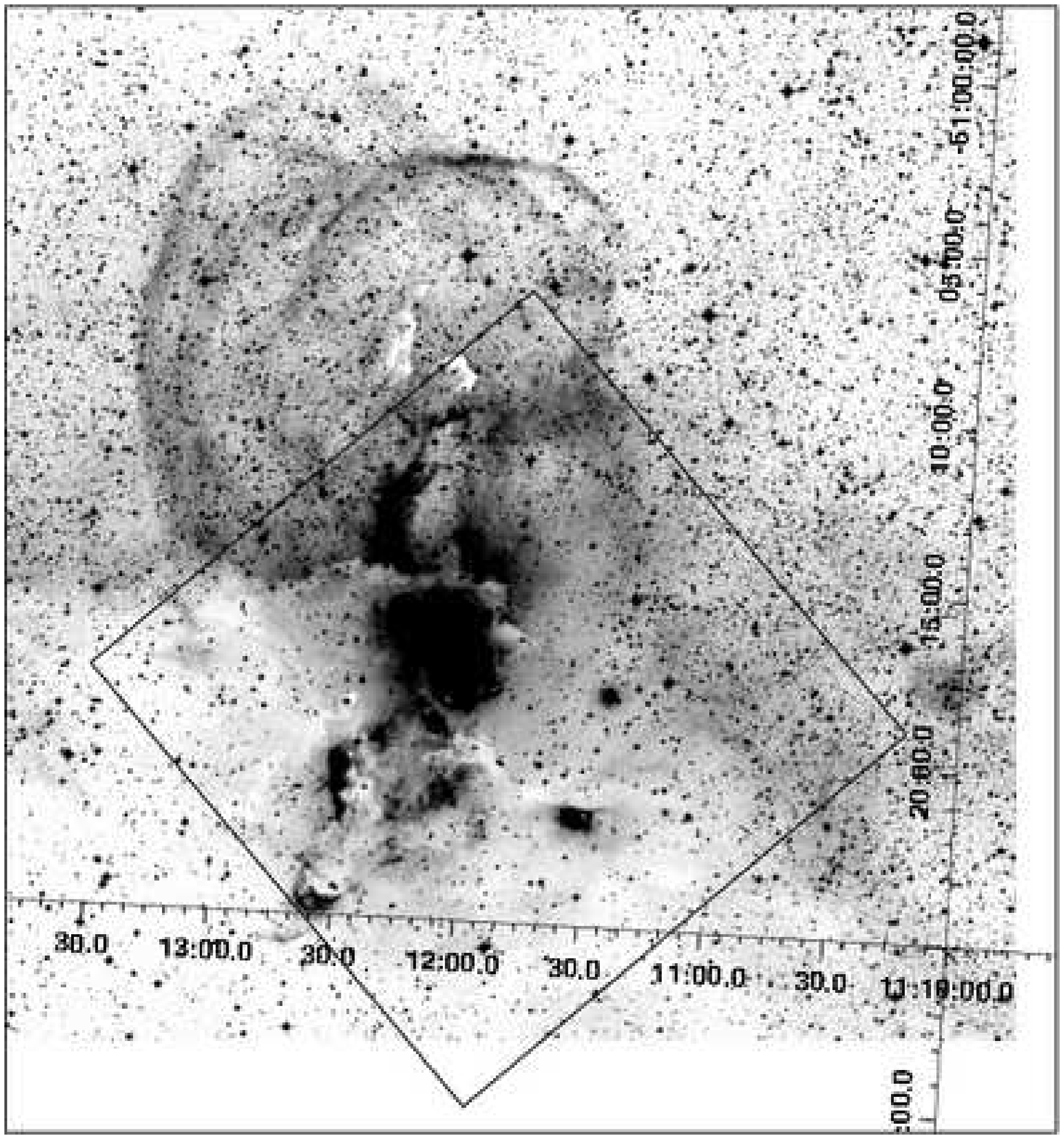,height=1.8in}
\hspace{0.2in}
      \psfig{figure=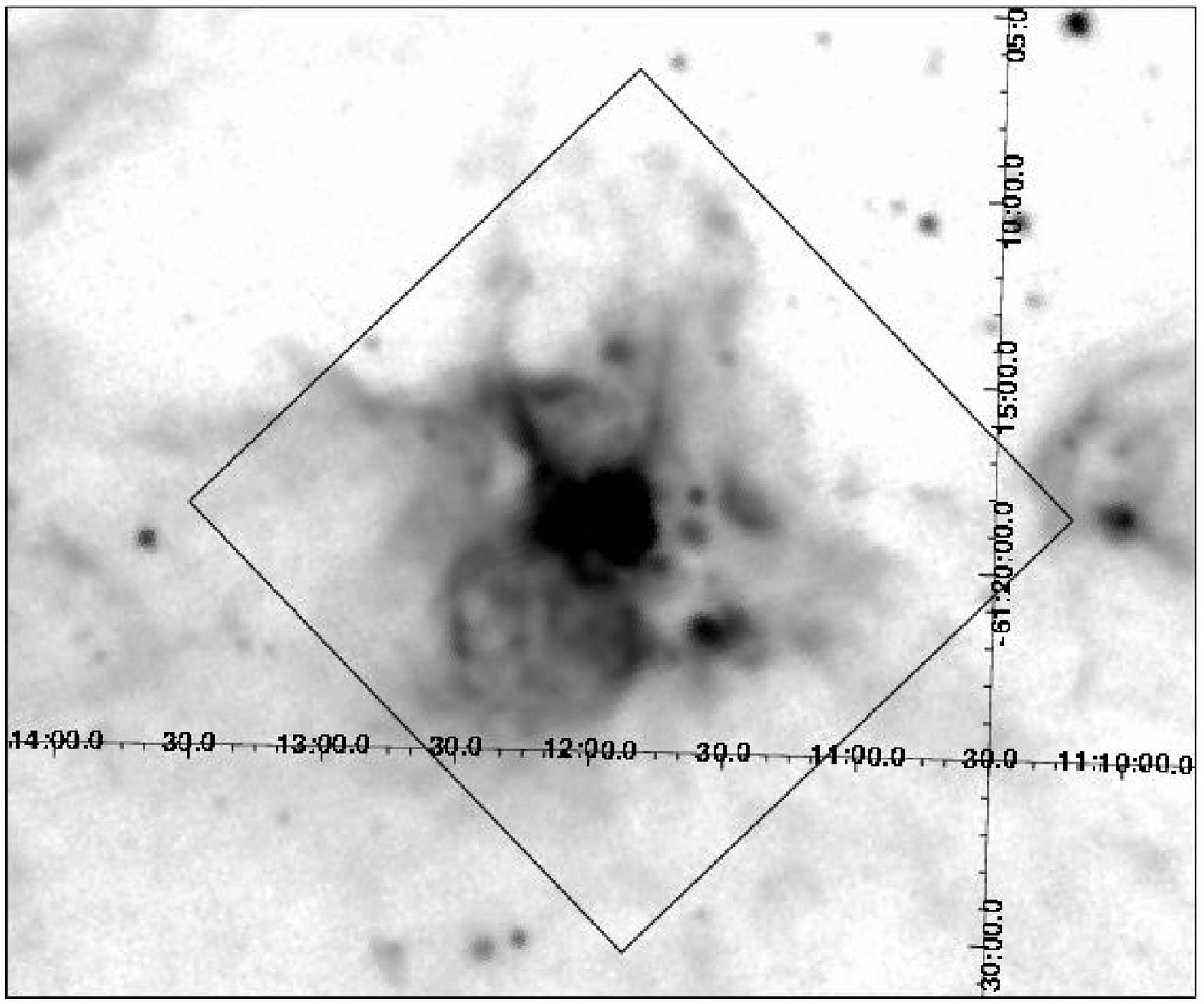,height=1.8in}}
\small
\caption{\protect \small NGC~3576 images.
{\bf Left:}  Digital Sky Survey red band.  
{\bf Right:} {\em MSX} 8$\mu$m.
The black squares show the size and approximate roll angle of the ACIS-I array.
}
\normalsize
\label{fig:3576_vis_ir}
\end{figure}

A radio study of NGC~3576 \citep{dePree99} revealed a large north-south
velocity gradient in the ionized gas of the nebula, indicating a
large-scale ionized outflow.  This flow may contribute to the large loops
and filaments seen in visual data (Figure~\ref{fig:3576_vis_ir}, left).
The 8$\mu$m {\em MSX} image (Figure~\ref{fig:3576_vis_ir}, right) is
complementary to the visual data, revealing heavy obscuration in the
southern half of the field and what appears to be a bipolar bubble likely
blown by the massive stars in NGC~3576's central cluster.  This bubble
appears closed to the south, where it encounters the dense GMC, but
open to the north, in the same direction as the large visual loops. 

In July 2005 we obtained a 60-ksec ACIS observation of NGC~3576, to see
if a soft X-ray bubble suggested by the {\em ROSAT} data was indicating
that NGC~3576 possessed an X-ray outflow similar to that seen in M17.
The ACIS aimpoint was directed at the strong infrared source IRS~1, at
the core of the embedded young stellar cluster, to search for embedded
protostars and the stellar sources responsible for ionizing NGC~3576, as
well as to resolve the brightest flaring pre-MS stars.  Our ACIS field
here is similar to that for M17:  only part of the bubble seen by {\em
ROSAT} is captured.  It was important to keep the aimpoint on IRS~1,
though, to study the embedded population and the IR cluster, which hold
the key to understanding the source of the {\em ROSAT} bubble.

\begin{figure} \centering

\mbox{\psfig{figure=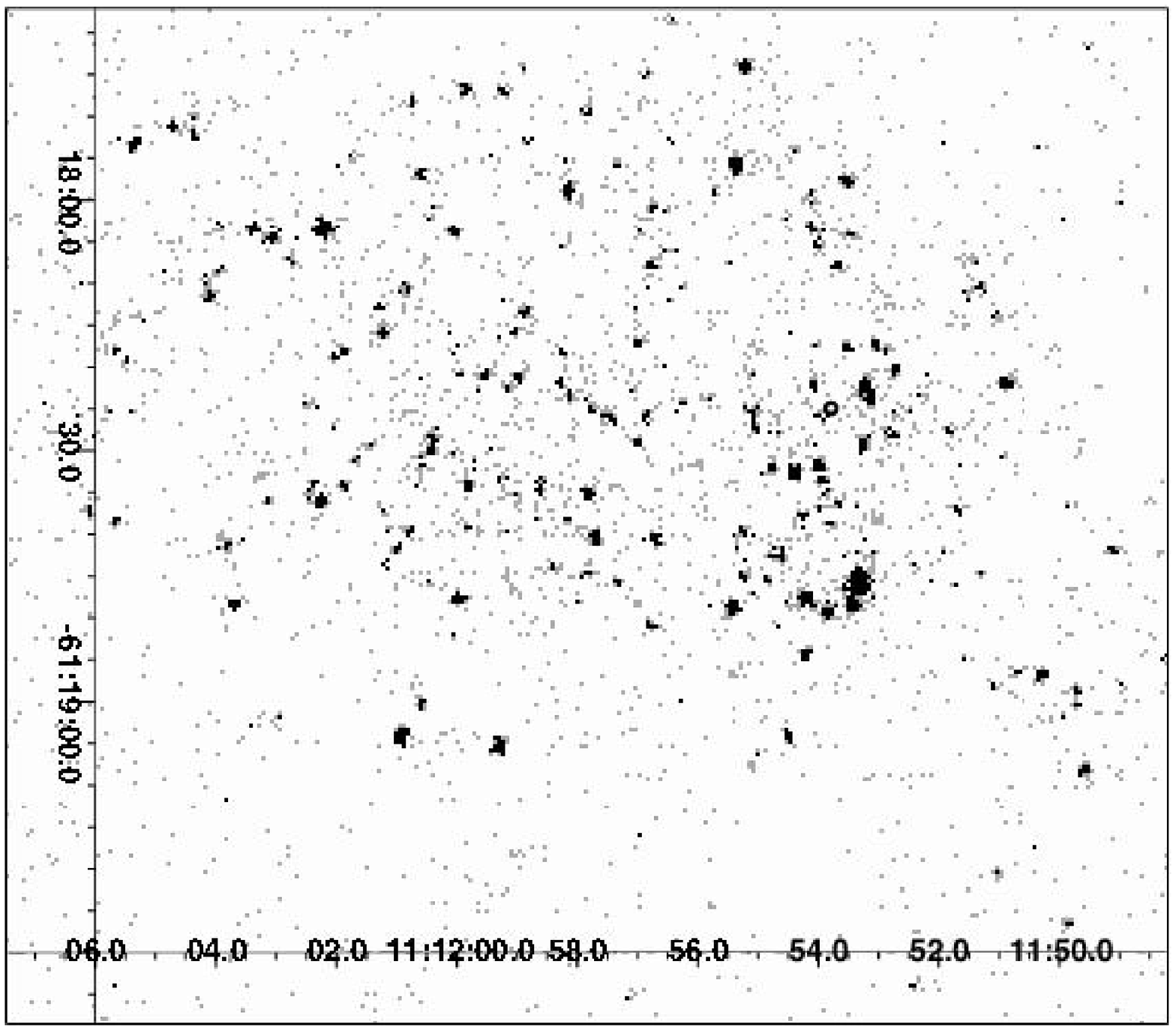,height=1.8in}
\hspace{0.2in}
      \psfig{figure=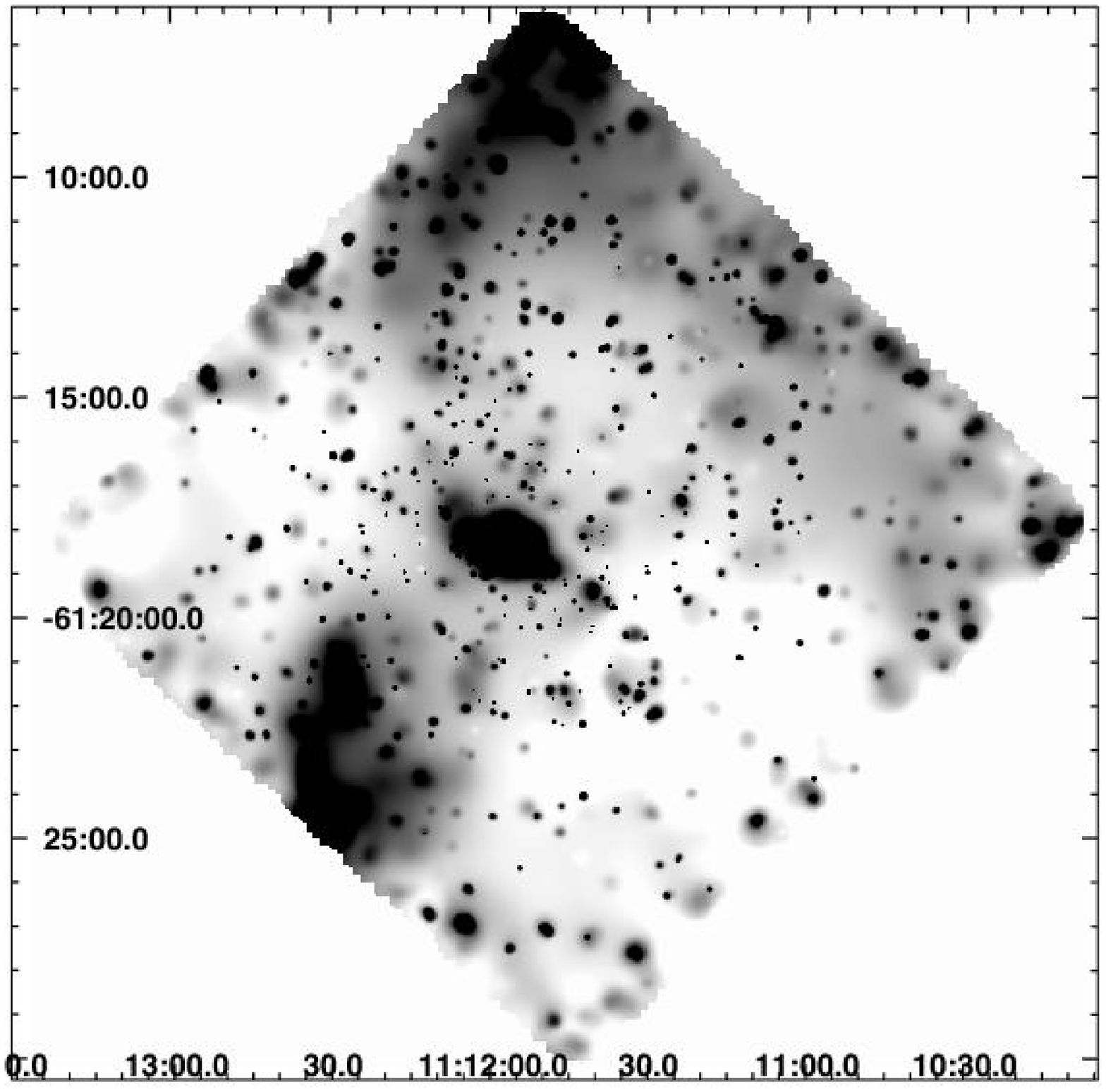,height=1.8in}}
\small
\caption{\protect \small NGC~3576 ACIS-I data.
{\bf Left:}  The central cluster.  The approximate position of IRS1 is
marked with a small black circle.
{\bf Right:}  Full-band (0.5--8 keV) smoothed image.
}
\normalsize
\label{fig:3576_acis}
\end{figure}

Figure~\ref{fig:3576_acis} (left) shows that this observation indeed
reveals the massive stellar engine powering NGC~3576, shredding the
molecular cloud from which it formed.  Many sources not known even
from mid-IR studies \citep[e.g.][]{Maercker06} are seen in X-rays.
The brightest of these is a hard, deeply-embedded source $\sim
20^{\prime\prime}$ south of IRS1.  This and other new X-ray sources may
be the long-sought embedded massive stars providing the extra ionization
for NGC~3576.

Much of the X-ray bubble seen by {\em ROSAT} is resolved
by {\em Chandra} into a wide array of point sources distributed across
the field.  Diffuse X-rays remain, however, near the top of the ACIS
field and in a more concentrated patch southeast of the central cluster.
Comparing the smoothed ACIS image (Figure~\ref{fig:3576_acis}, right) to
the visual and mid-IR data in Figure~\ref{fig:3576_vis_ir}, it appears
that the northern visual loops outline (at least in part) soft diffuse
X-ray emission in and beyond the northern, more open lobe of the bipolar bubble.
Throughout the field the diffuse X-ray emission is complementary to the
mid-IR emission, suggesting either that warm dust has displaced the hot
X-ray gas or that this material is shadowing the X-rays, absorbing and
obscuring any soft X-ray emission that may lie behind it.  The southern
lobe is also traced by soft diffuse X-ray gas; the lack of shadowing here
implies that this part of the wind-blown bubble lies on the near side of
the GMC.  There are clear relationships between NGC~3576's X-ray plasma,
velocity and temperature gradients measured in the radio, the impressive
visual loops, and the mid-IR bipolar bubble.
Spectral analysis of this diffuse X-ray emission is ongoing and should
give some insight into the energetics of this important complex.

{\em Chandra} has revealed that NGC~3576 and M17 are two examples of
the same phenomenon, although NGC~3576's cluster is younger and more
deeply embedded in its natal cloud, leading to the more complex spatial
morphology in its soft diffuse X-ray emission.  It appears that
hot, flowing X-ray gas from OB winds may be a common component of young
blister HII regions, strongly affecting their morphology and dynamics.

\section{W3:  Ionizing Sources Revealed and a Missing Cluster}

W3 is a large, nearby \citep[$D=2.0$~kpc,][]{Hachisuka06} MSFR in
the Perseus Arm of the Milky Way.  The W3 complex contains one of the
most massive molecular clouds in the outer Galaxy \citep{Heyer98},
massive embedded protostars \citep{Megeath05}, near- and mid-IR sources
\citep{Kraemer03}, masers, and outflows.  It is unique in containing all
morphological classes of HII region, from hypercompact to diffuse, 0.01
to 1~pc in diameter, with ages $10^3$--$10^6$~yrs \citep{Tieftrunk97}.

Many authors have argued that star formation in W3 is being induced
by the expansion of the adjacent W4 superbubble, which is sweeping up
molecular gas into a high-density layer, within which stars are forming.
\citet{Oey05} revised this scenario, proposing that the young (3--5
Myr) OB cluster IC~1795, triggered to form by W4, is blowing its own
second-generation superbubble at the molecular cloud interface, triggering
in turn the W3 complex of massive star formation.  W3 is an ideal
testbed for understanding recent, ongoing, and triggered star formation;
X-ray sources there are all at the same distance yet exist in a range of
dynamical settings, providing a series of controlled environments in which
to study the violent interactions of massive stars with their natal cloud.

\begin{figure} \centering

\psfig{figure=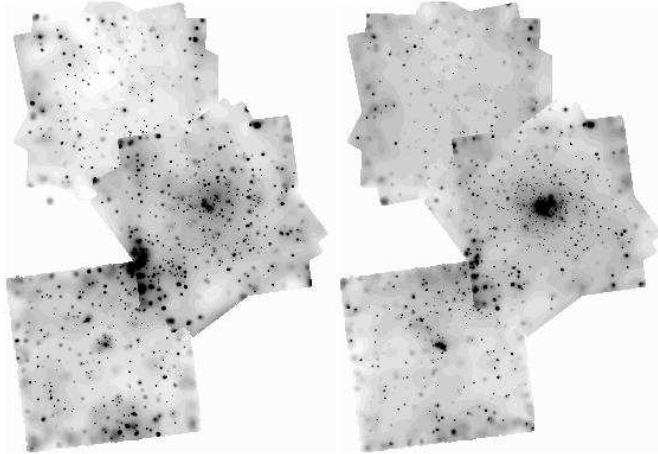,height=2.4in}
\small
\caption{\protect \small {\bf Left:}  Smoothed soft-band (0.5--2~keV)
ACIS-I mosaic of W3 North (top), W3 Main (middle), and W3(OH) (bottom).
The slightly older OB cluster IC~1795 is partially imaged between W3
Main and W3(OH).
{\bf Right:}  Hard-band (2--8~keV) mosaic of the same W3 regions.
}
\normalsize
\label{fig:w3_soft-hard}
\end{figure}

With many observations throughout 2005, used ACIS-I to image the three
major star-forming complexes, W3 North, W3 Main, and W3(OH), amassing
$\sim 80$~ksec on each field.  A mosaic of these observations is shown
in Figure~\ref{fig:w3_soft-hard}.  ACIS resolves the OB stars powering
the HII regions, embedded massive stars and protostellar objects, and
the pre-MS population, yielding $>1300$ X-ray sources in the three $\sim
80$~ksec pointings.

W3~North is a well-developed parsec-scale HII region powered by the
O6V star GSC 04050-02567 that may be isolated, with no lower-mass
accompanying population \citep{Carpenter00}.  The HII region is so bright
in the near-IR due to nebular emission that it is difficult to perform
an IR search for an underlying cluster.  {\em Chandra} detects the O6
star; it is a modest, soft X-ray emitter, typical of single O stars.
However Figure~\ref{fig:w3_soft-hard} clearly reveals that W3 North
is not an Orion-like field; no cluster of young X-ray-emitting pre-MS
stars is seen.  This may indicate that the region has an anomalous IMF,
that this O6 star is a young example of a massive star that formed in
isolation \citep{deWit05}, or that the O6 star is a runaway from either
IC~1795 or W3~Main that has somehow managed to create a large and complex
HII region at the edge of W3's high-density layer.  In any case, this
is an unusual situation that merits more detailed study.

W3~Main harbors the strongest CO peak in W3 and IR clusters with over 200
stars \citep{Carpenter00}.  An early 40-ksec ACIS-I observation showed
possible extended emission spatially coincident with the HII regions;
over 100 of the 236 ACIS sources detected in this original dataset are
near the W3 core \citep{Hofner02}.  We added another 40 ksec to this
original observation; in the composite image, it is clear that W3 Main
is the dominant star-forming center of W3, with an extensive population
of young stars nearly filling the $17^\prime \times 17^\prime$ ACIS-I
field of view.  The full extent of W3 Main was not clear from earlier
IR studies.  Of the $>1300$ sources in the W3 ACIS mosaic, over half of
them are captured in the W3 Main pointing.

W3(OH) contains the B0.5 star IRAS~02232+6138 surrounded by a
cluster of more than 200 stars \citep{Carpenter00}, an UCHII region
(2$^{\prime\prime}$ in diameter), OH and H$_2$O masers, outflows,
IR sources, and strong CO indicating massive embedded protostars.
ACIS detects a point source associated with the W3(OH) UCHII region
and resolves the string of known IR stellar clusters northeast of
W3(OH) that may indicate a broader region of ongoing star formation
\citep{Tieftrunk97}.
  
\begin{figure} \centering

\psfig{figure=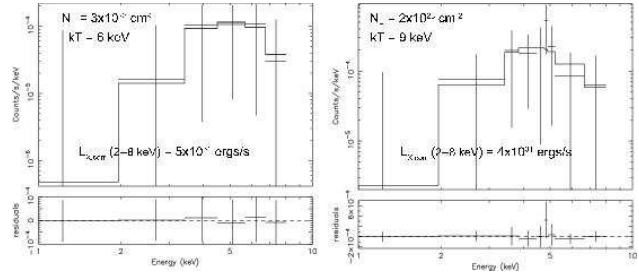,height=1.4in}
\small
\caption{\protect \small ACIS spectra of embedded W3 protostars.  
{\bf Left:}  W3 Main IRS5.
{\bf Right:}  W3(OH).
}
\normalsize
\label{fig:w3_spectra}
\end{figure}
  
Figure~\ref{fig:w3_spectra} shows the ACIS spectra of two interesting
embedded sources in the W3 complex:  IRS5 in W3 Main and the X-ray source
in W3(OH), presumably the source of ionization for the UCHII region.
IRS5 is thought to be a multiple system of perhaps five proto-OB stars
\citep{Megeath05}.  We don't resolve the components with ACIS, but
we do find a very hard spectrum for the X-ray source coincident with
IRS5, again consistent with colliding-wind binary emission, although
this hard emission could be coming from a cluster wind generated by
the combined effects of the winds from all of these massive protostars
\citep{Canto00}.  The W3(OH) source is also extremely hard.  Due to the
high obscuration toward these sources, we have no information on any soft
X-ray emission that they might be generating.  The luminosities that we
note in Figure~\ref{fig:w3_spectra} are corrected for this absorption,
but are only given for the hard (2--8~keV) X-ray band.

\section{Trumpler 14:  Swarms of Stars and Misplaced Hot Gas}

The Carina complex is part of the Sagittarius-Carina spiral arm, at
a distance of $\sim$2.3~kpc \citep{Smith06b}.  It is a remarkably rich
star-forming region, containing 8 open clusters with at least 66 O stars,
several Wolf-Rayet (WR) stars, and the luminous blue variable Eta Carinae
\citep{Smith06a}.  The combined Carina OB clusters Tr16, Tr14, and Cr228
contain the nearest rich concentration of early O stars; their ionizing
flux and winds may be fueling a young bipolar superbubble \citep{Smith00}.
High ISM velocities throughout the complex \citep{Walborn02b} and the
presence of evolved stars may imply that a supernova might already
have occurred in this region, although no well-defined remnant has ever
been seen. 

Tr14 is an extremely rich, young ($\sim$1~My), compact OB cluster near
the center of the Carina complex, containing at least 30 O and early
B stars \citep{Vazquez96}.  The radio HII region Carina I is situated
just west of Tr14, at the edge of the GMC and near a strong CO peak.
It is ionized by Tr14 and is carving out a cavity in the molecular
cloud which now contains {\em IRAS} sources, high-mass protostars,
their associated UCHII regions, and ionization fronts \citep{Brooks01}.
Although star formation has ceased in the Keyhole region of Carina due to
the harsh environment created by the massive stars there, the proximity
of Tr14 to the GMC has triggered a new generation of star formation in
Carina I \citep{Rathborne02}.
 
\begin{figure} \centering

\mbox{\psfig{figure=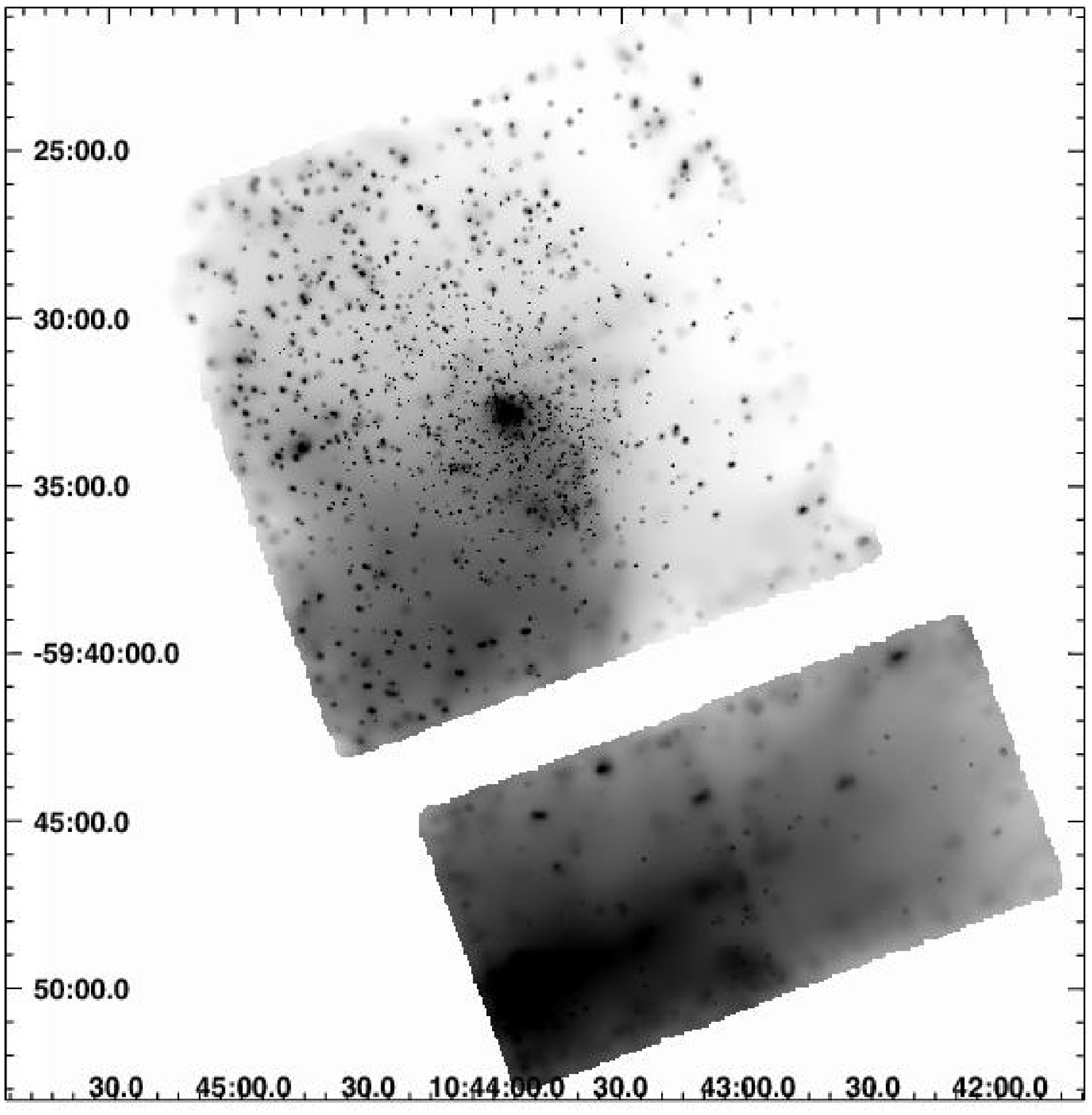,height=1.8in}
\hspace{0.2in}
      \psfig{figure=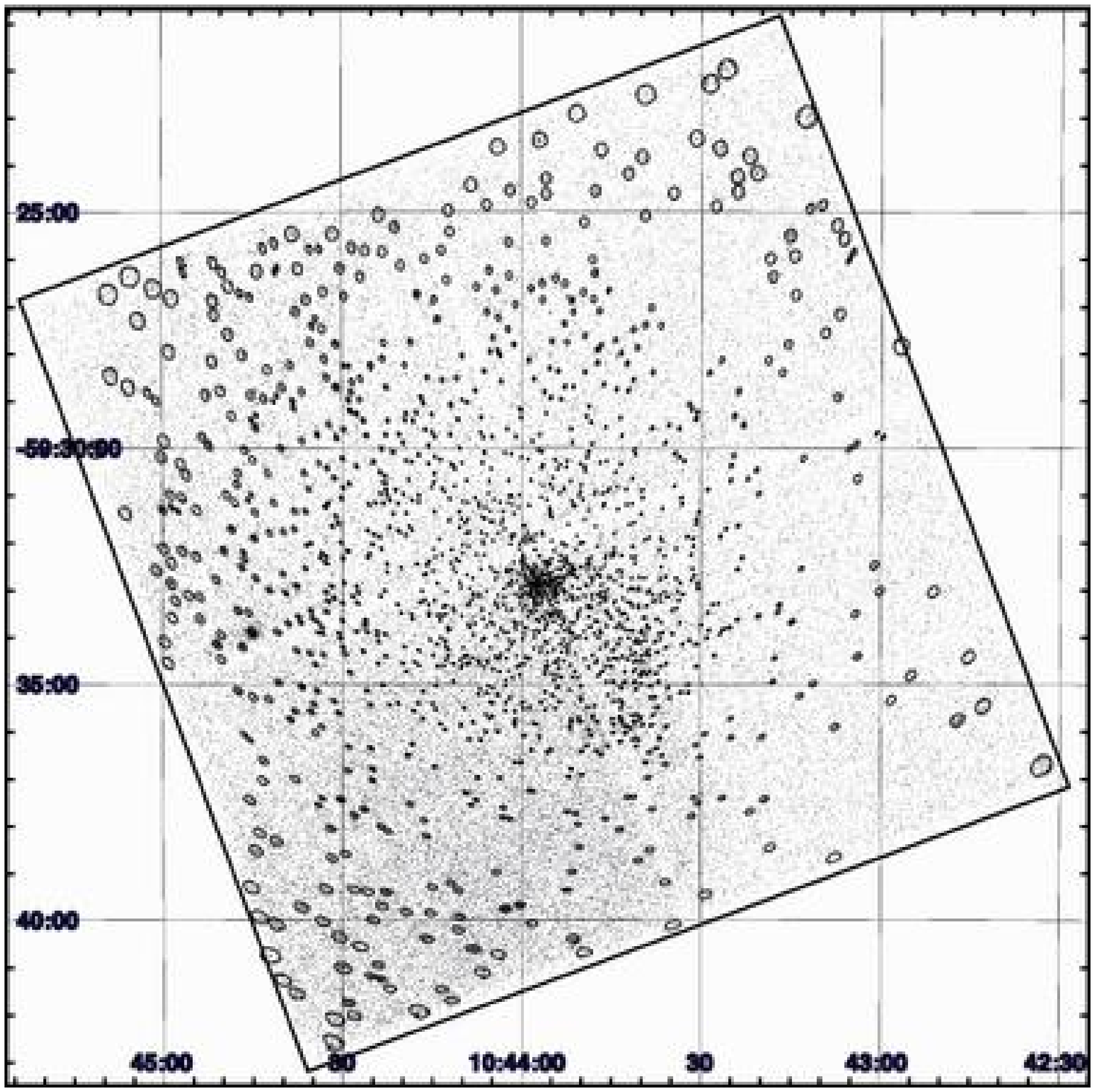,height=1.8in}}
\small
\caption{\protect \small ACIS images of Tr14 in Carina.
{\bf Left:}  Full-field, full-band smoothed image.  
{\bf Right:}   The ACIS-I image (binned by $2^{\prime\prime}$), showing
$\sim$1600 point source extraction regions.
}
\normalsize
\label{fig:tr14_images}
\end{figure}

In September 2004 we obtained a 57-ksec ACIS-I observation of Tr14
(Figure~\ref{fig:tr14_images}).  The aimpoint of this observation is the
central star in the cluster, HD~93129AB, a very early-type (O2I--O3.5V)
binary \citep{Walborn02a}, with the two components separated by $\sim
3^{\prime\prime}$.  Our ACIS-I observation reveals $\sim$1600 X-ray point
sources in the Tr14 region, suffused by bright, soft diffuse emission.
Since the Tr16 and Tr14 clusters overlap, some sources toward the east
and southeast are most likely Tr16 members; a few sources in the northeast
corner of our field may be members of the nearby Tr15 cluster.  The soft
diffuse emission pervading the eastern half of our field is sharply cut
off to the west by the GMC.  The density of X-ray point sources also
falls sharply to the west, due in part to reduced sensitivity because of
higher extinction and in part because star formation is just now being
triggered in that region, as described above. 

\begin{figure} \centering

\mbox{\psfig{figure=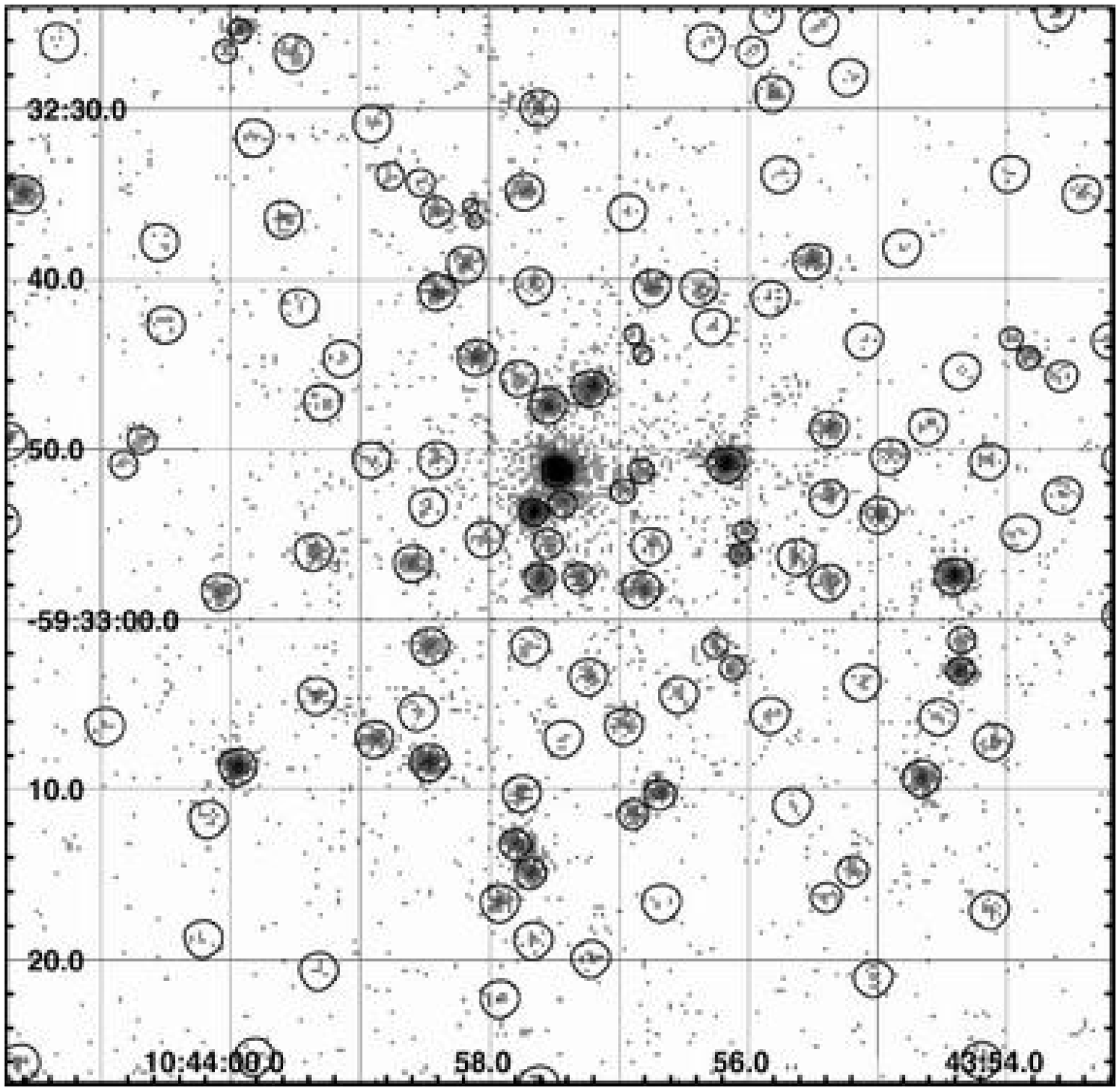,height=1.8in}
      \psfig{figure=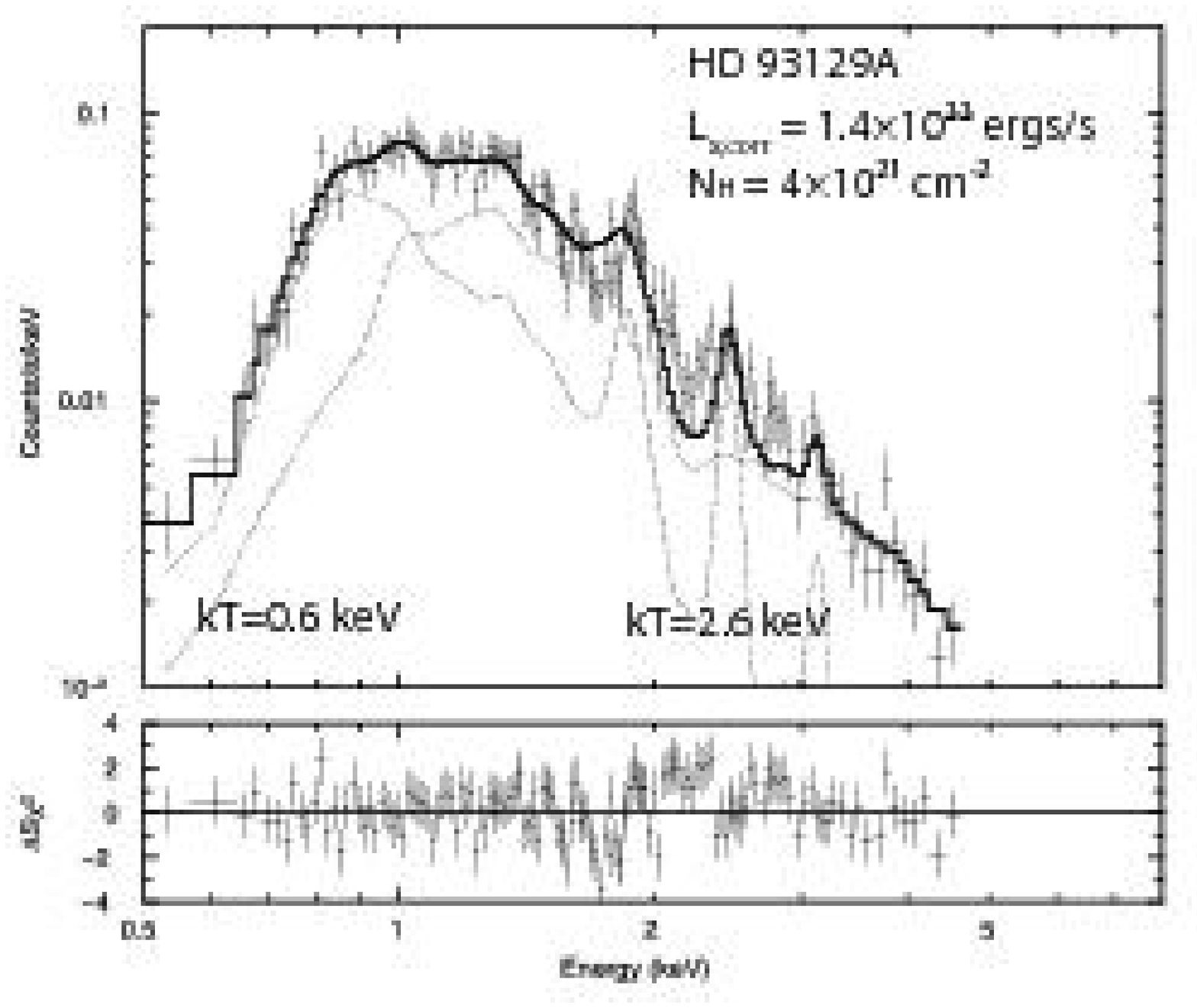,height=1.5in}
\hspace*{-0.2in}
      \psfig{figure=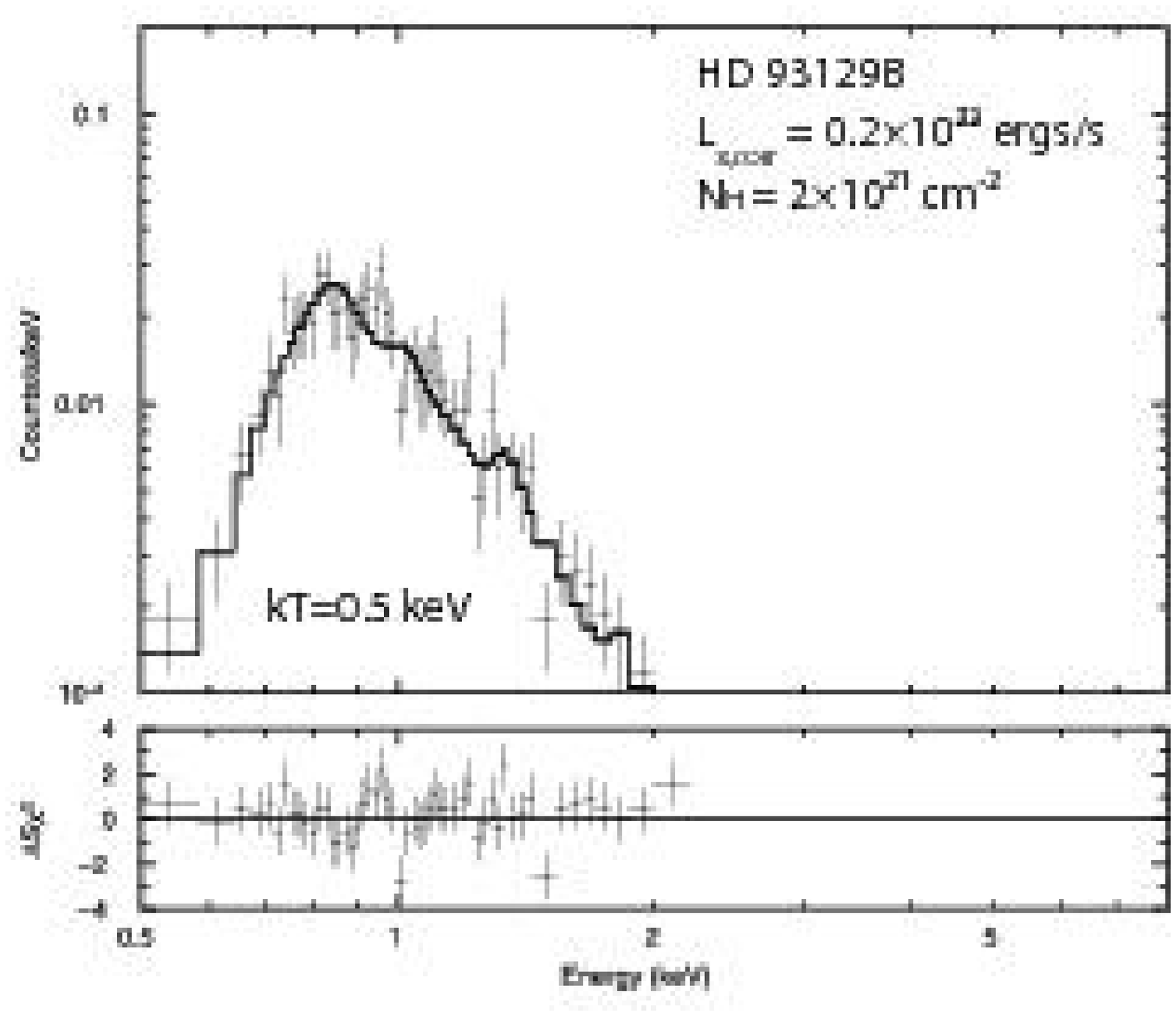,height=1.5in}}
\small
\caption{\protect \small {\bf Left:}  The central part of the ACIS-I observation of Tr14 with extraction regions marked.
{\bf Center:} ACIS spectrum of the resolved massive binary at the aimpoint
of the Tr14 observation, HD93129A (O2If\*) and
{\bf Right:} HD93129B (O3V).
}
\normalsize
\label{fig:tr14_ptsrcs}
\end{figure}

Figure~\ref{fig:tr14_ptsrcs} (left) shows the crowded central region of
Tr14 in X-rays, with the source extraction regions drawn.  The brightest
source near field center is HD93129A; it is clearly resolved from
HD93129B (seen slightly to the southeast) and from another, fainter
star just to its south (west of HD93129B).  The ACIS spectra of these
early O stars are shown in Figure~\ref{fig:tr14_ptsrcs} (right panel).
Although these two stars are of similar early spectral type, they show
remarkably different X-ray spectra.  While both exhibit the expected soft
thermal plasma with $kT \sim 0.5$~keV, HD93129A also requires a second,
harder thermal component and is 7 times brighter in X-rays than HD93129B.
We find similar results for other O stars in the field:  while the O3V
star HD93128 is soft ($kT = 0.3$~keV) and faint (absorption-corrected
luminosity $L_{\rm x,corr} = 0.1 \times 10^{33}$~ergs/s), the O3V star
HD93250 is the brightest source in the ACIS field, with  $L_{\rm x,corr}
= 2.1 \times 10^{33}$~ergs/s and requiring both soft ($kT = 0.6$~keV) and
hard ($kT = 3.3$~keV) thermal plasma components.  As for other sources
described above, this leads us to speculate that HD93129A and HD93250
could be colliding-wind binaries; alternatively the hard components of
their spectra are soft enough that they could be magnetically active
early O stars, similar to $\theta^1$~Ori~C \citep{Gagne05}.  In contrast,
HD93129B and HD93128 are likely more ``normal'' O stars without these
phenomena.  

\begin{figure} \centering

\mbox{\psfig{figure=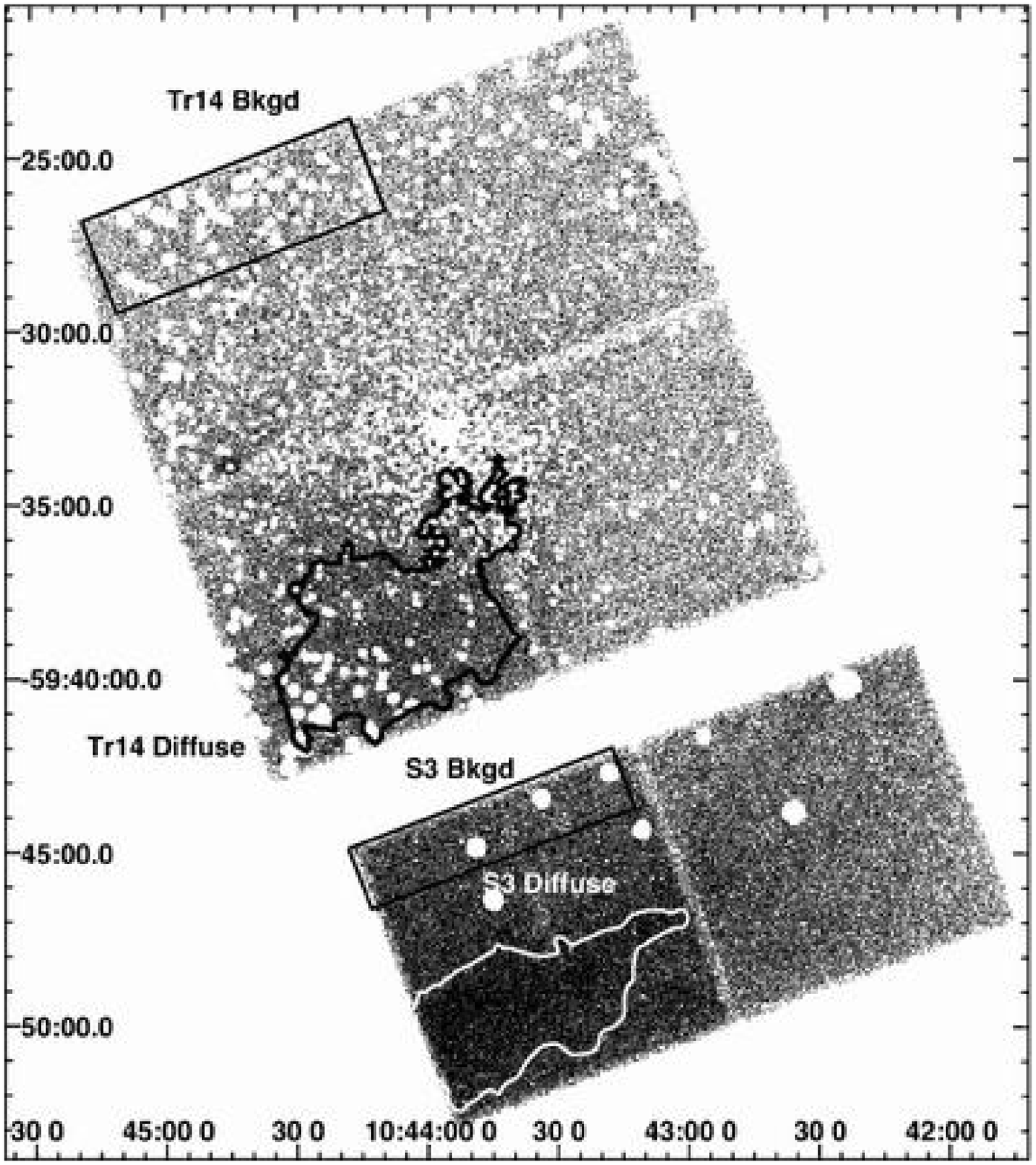,height=1.8in}
      \psfig{figure=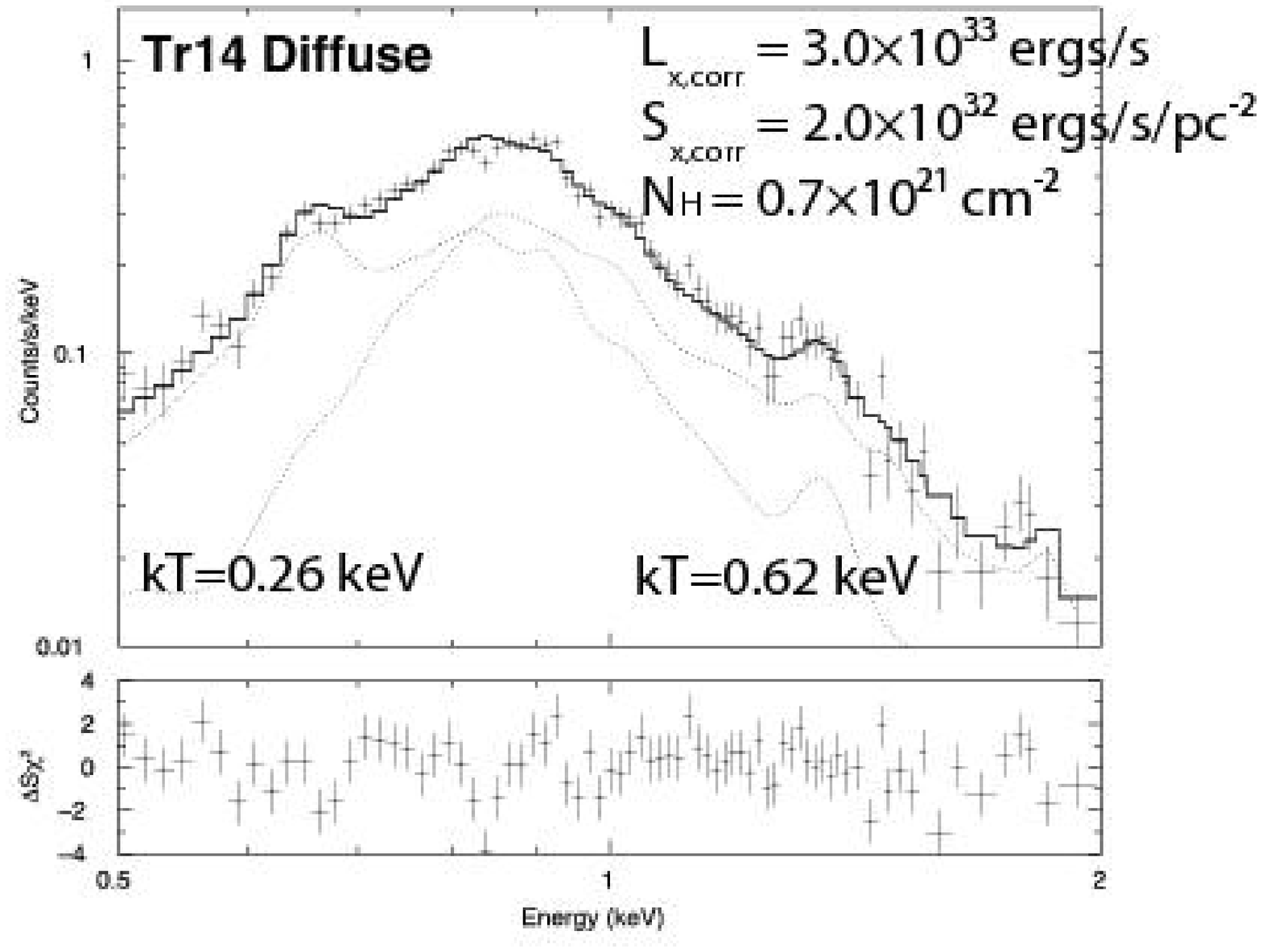,height=1.5in}
\hspace*{-0.2in}
      \psfig{figure=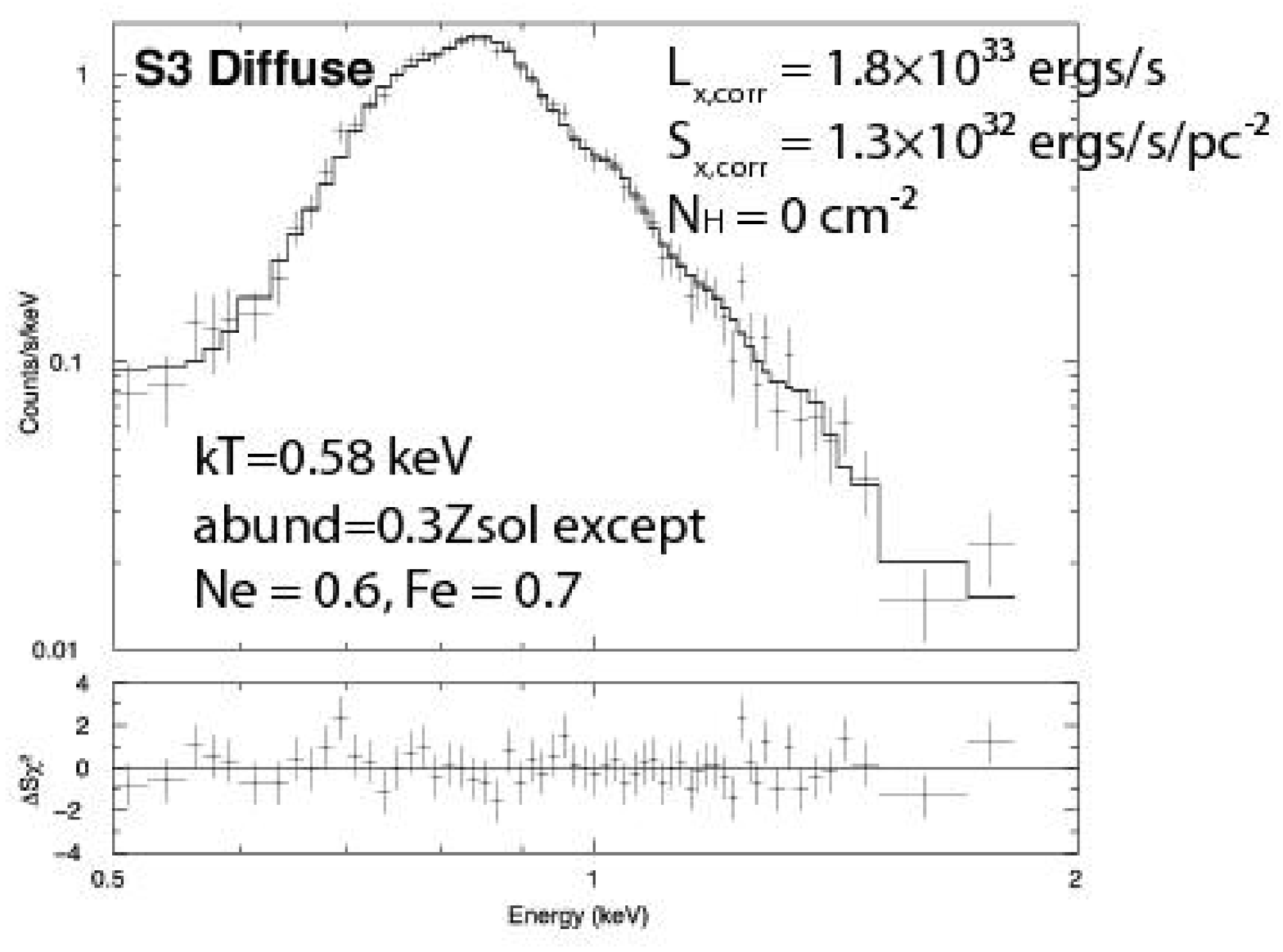,height=1.5in}}
\small
\caption{\protect \small Diffuse emission in Tr14.
{\bf Left:}  Image showing point sources removed, with diffuse extraction
regions marked.
{\bf Center and Right:}  Spectra of the two diffuse regions.
}
\normalsize
\label{fig:tr14_diffuse}
\end{figure}

The soft diffuse emission seen in this ACIS observation of Tr14
(Figure~\ref{fig:tr14_diffuse}, left) is also remarkable.  On the
ACIS-I array, this diffuse emission is brightest in the southeast
quadrant, in the region between Tr14 and Tr16.  Surprisingly, it is not
centered on the Tr14 cluster.  The apparent surface brightness of this
emission may be affected by gradients in the absorbing column across
the field.  We serendipitously imaged very bright diffuse emission in
the off-axis ACIS-S CCDs that were also operational for this observation.
This emission is far from the known massive stars in the Carina complex.
Although its apparent surface brightness is much higher than the diffuse
emission seen on the ACIS-I array, it is intrinsically fainter; we see
it as brighter because it suffers no measurable absorption.  This may
indicate that it lies in front of the Tr14 cluster, perhaps partially
filling the lower lobe of the Carina superbubble seen in the mid-IR.  

While both regions of diffuse emission exhibit soft spectra
(Figure~\ref{fig:tr14_diffuse}), they require quite different
model fits.  The emission in the southeast quadrant of the I array
requires a two-temperature fit ($kT = 0.3$~keV and $kT = 0.6$~keV),
some absorption, and typical low abundances of $Z = 0.3Z_{\odot}$.  The
S array emission requires just a single thermal component with $kT =
0.6$~keV, but no absorbing column is seen and a good fit is only
obtained by increasing the abundances of Ne and Fe, possibly indicating
a supernova origin for this emission.  Earlier X-ray data suggested
that the Carina complex is pervaded by diffuse X-ray emission not
concentrated on the star clusters.  This detailed view from {\em
Chandra} now shows that the diffuse emission is spatially and
spectrally complex and, while it may be due in part to massive star
winds, one or more cavity supernovae probably also contribute to its
complexity.  More high-resolution observations of the Carina MSFR are
necessary to untangle this complicated mix of pointlike and diffuse
X-ray sources.

\section{30 Doradus:  Live Fast, Die Young, Blow Some Bubbles}

30 Doradus in the Large Magellanic Cloud is the most luminous Giant
Extragalactic HII Region and ``starburst cluster'' in the Local Group.
At the center of 30~Dor is the ``super star cluster'' R136,  with dozens
of 1-2 Myr-old $> 50$ M$_\odot$ O and Wolf-Rayet stars \citep{Massey98}.
30~Dor's superbubbles are well-known bright X-ray sources, where multiple
cavity supernovae from past OB stars produce soft X-rays filling bubbles
50--100~pc in size with $L_x \sim 10^{35-36}$~ergs/s \citep{Chu90}.

\begin{figure} \centering

\psfig{figure=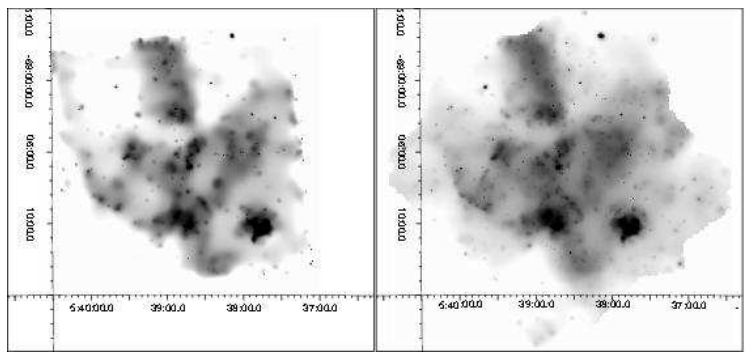,height=1.8in}
\small
\caption{\protect \small {\em Chandra}/ACIS-I soft-band images of 30~Doradus.
{\bf Left:}  The original 23-ksec observation.  
{\bf Right:}  New data combined with the original observation, making
a $\sim 110$~ksec image.
}
\normalsize
\label{fig:30dor_new-old}
\end{figure}

Figure~\ref{fig:30dor_new-old} (left) shows the smoothed ACIS-I
image from our original 23-ksec observation of 30~Dor, described in
\citet{Townsley06a,Townsley06b}.  Using super-resolution techniques,
we find $\sim 100$ point sources in R136, but clearly the dominant
X-ray structures in 30~Dor remain the plasma-filled superbubbles.
We constructed maps of the diffuse X-ray emission in those superbubbles,
showing variations in plasma temperature (T = 3--9 million
degrees), absorption ($N_H$ = 1--6$\times 10^{21}$~cm$^{-2}$), and
absorption-corrected X-ray surface brightness ($S_{\rm x,corr}$ = 3--126$\times
10^{31}$~ergs~s$^{-1}$~pc$^{-2}$).  Some new X-ray concentrations
$\simeq 30^{\prime\prime}$ (7 pc) in extent are spatially associated
with high-velocity optical emission line clouds \citep{Chu94}.
Figure~\ref{fig:30dor_new-old} (right) shows the original short ACIS-I
observation combined with a new 90-ksec dataset obtained in January 2006.
As expected, more point sources are resolved in the longer observation.
Smaller-scale diffuse structures are also emerging, likely tracing more
subtle shock features across the complex.  No obvious hard diffuse
emission from a cluster wind in R136 is seen, even in this longer
observation.

\begin{figure} \centering

\psfig{figure=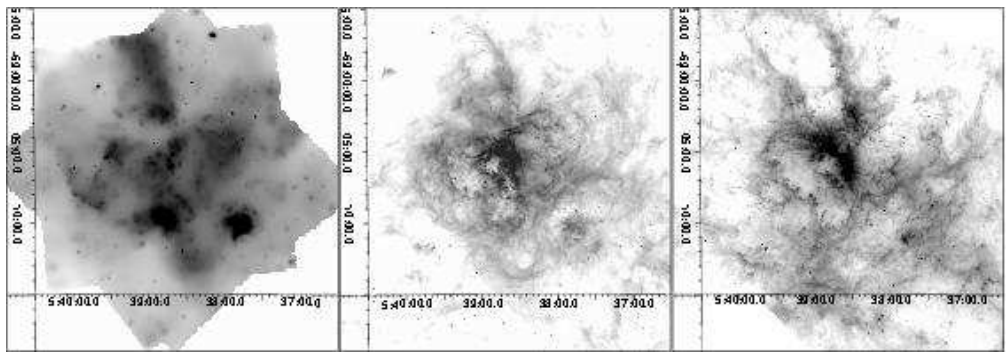, height=1.8in}
\small
\caption{\protect \small  30~Doradus across the spectrum.
{\bf Left:}  The new soft-band ACIS image.
{\bf Center:} An H$\alpha$ image from MCELS.
{\bf Right:} The {\em Spitzer} 8$\mu$m image.
}
\normalsize
\label{fig:30dor_3bands}
\end{figure}

Figure~\ref{fig:30dor_3bands} shows the deeper ACIS image in
context with visual and IR data.  The center panel shows the recently
released H$\alpha$ image from the Magellanic Cloud Emission Line Survey
\citep[MCELS,][]{CSmith00}, while the right panel shows the 8$\mu$m {\em
Spitzer}/IRAC image ({\em Spitzer} press release by B.\ Brandl, 2004).
The MCELS and {\em Spitzer} data gives new insight into the diffuse X-ray
structures:  the hot superbubbles fill the interiors of ionization fronts
outlined by H$\alpha$ emission, which in turn are outlined by shells of
warm dust; the X-ray confinement and morphology are fully appreciated
only when anchored by these multiwavelength data.

Our spectral analysis of the superbubbles seen in the early ACIS data
revealed a range of absorptions, plasma temperatures, and abundance
variations.  This short observation limits our ability to study diffuse
structures at small scales; in order to get enough counts to do good
spectral fitting, we must average over many tens of parsecs.  We suspect
that we are thus averaging over many distinct plasma components, with
different pressures, temperatures, absorbing columns, and possibly even
different abundances.  The new observation will allow us to refine our
spectral fitting and to make higher-resolution maps of the physical
parameters that govern 30~Dor's diffuse X-ray emission.  We will also be
able to study more completely the massive stars in R136 and throughout
the complex.

\section{Summary}

High-resolution X-ray studies of MSFRs reveal
magnetically-active pre-MS stars, massive star microshocks, and
colliding-wind binaries.  {\em Chandra} provides a disk-unbiased stellar
sample for IR study.  Diffuse X-ray emission is also seen, caused by O
star winds interacting with themselves and with the surrounding media.
Supernovae occurring in the cavities blown by OB clusters produce brighter
soft X-rays, dominating the X-ray emission in large multi-cluster and
multi-generation star-forming complexes.

Several {\em Chandra} discoveries are noted in this short overview.
For all regions, early O stars often show anomalously hard spectra,
perhaps indicating close binarity.  For individual regions, discoveries include:\\
$\bullet$ M17:	First clear detection of an X-ray outflow.\\
$\bullet$ NGC 3576:  Bipolar bubble seen in {\em MSX} data is likely
the wind-blown bubble from an embedded OB cluster; this field shows the
second clear example of X-ray outflowing gas; hard X-rays reveal cluster
members not seen in the IR -- these are likely the missing ionizing
massive stars.\\
$\bullet$ W3:  W3 Main is huge compared to other W3 young clusters;
the W3 Main IRS5 and W3(OH) massive protostars are very
hard X-ray sources; W3 North is missing its cluster.\\
$\bullet$ Tr14:  Some early O stars are likely binary or magnetically active, some are not; soft
diffuse emission appears between Tr14 and Tr16, while separate, unabsorbed
soft diffuse emission located far from massive stars may indicate a cavity supernova remnant inside the Carina complex's young superbubble.\\
$\bullet$ 30 Dor:  More spatial structure and point sources emerge in
the longer observation; no clear sign of hard X-rays from a cluster wind
in R136 are seen.

\begin{acknowledgments}
I am most indebted to my colleagues Patrick Broos for data analysis
and figure production, Eric Feigelson for ideas and interpretation,
and Gordon Garmire, the ACIS PI, for using part of his {\em Chandra}
guaranteed time on the MSFR project.  I thank Barbara Whitney and Ed
Churchwell for supplying the GLIMPSE image of M17, You-Hua Chu for the
MCELS image of 30~Dor, and Bernhard Brandl for the IRAC image of 30~Dor.
Support for this work was provided by NASA through contract NAS8-38252
and {\em Chandra} Awards GO4-5006X, GO4-5007X, GO5-6080X, GO5-6143X,
and SV4-74018 issued by the Chandra X-ray Observatory Center, operated
by the Smithsonian Astrophysical Observatory for and on behalf of NASA
under contract NAS8-03060. 
\end{acknowledgments}


\begin{thebibliography}{}

\bibitem[Arnaud(1996)]{Arnaud96} \textsc{Arnaud, K.~A.} 1996  {XSPEC: The First Ten Years.} In \textit{Astronomical Data Analysis Software and Systems V} (ed.\ G.\ Jacoby \& J.\ Barnes),  ASP Conf.~Ser.\ \textbf{101}, 17

\bibitem[Brooks et al.(2001)]{Brooks01} \textsc{Brooks, K.~J., Storey, 
J.~W.~V., \& Whiteoak, J.~B.} 2001  {H110$\alpha$ recombination-line emission and 4.8-GHz continuum emission in the Carina nebula.} \mnras, \textbf{327}, 46 

\bibitem[Broos et al.(2006)]{Broos06} \textsc{Broos, P.~S., Getman, K.~V., Townsley, L.~K., Feigelson, E.~D., Wang, J., Garmire, G.~P., Jiang, Z., \& Tsuboi, Y.} 2006  {The Young Stellar Population in M17 Revealed by Chandra.} \apj, submitted

\bibitem[Cant{\' o} et al.(2000)]{Canto00} \textsc{Cant{\' o}, J., 
Raga, A.~C., \& Rodr{\'{\i}}guez, L.~F.} 2000  {The Hot, Diffuse Gas in a Dense Cluster of Massive Stars.} \apj, \textbf{536}, 896 
 
\bibitem[Carpenter et al.(2000)]{Carpenter00} \textsc{Carpenter, J.~M., 
Heyer, M.~H., \& Snell, R.~L.} 2000  {Embedded Stellar Clusters in the W3/W4/W5 Molecular Cloud Complex.} \apjs, \textbf{130}, 381 

\bibitem[Chu \& Mac Low(1990)]{Chu90} \textsc{Chu, Y.~\& Mac Low, M.} 1990  {X-rays from superbubbles in the Large Magellanic Cloud.}
\apj, \textbf{365}, 510

\bibitem[Chu \& Kennicutt(1994)]{Chu94} \textsc{Chu, Y.-H., \& 
Kennicutt, R.~C., Jr.} 1994  {Kinematic structure of the 30 Doradus giant H II region.} \apj, \textbf{425}, 720 

\bibitem[de Pree et al.(1999)]{dePree99} \textsc{de Pree, C.~G., 
Nysewander, M.~C., \& Goss, W.~M.} 1999  {NGC 3576 and NGC 3603: Two Luminous Southern H II Regions Observed at High Resolution with the Australia Telescope Compact Array.} \aj, \textbf{117}, 2902 

\bibitem[de Wit et al.(2005)]{deWit05} \textsc{de Wit, W.~J., Testi, 
L., Palla, F., \& Zinnecker, H.} 2005  {The origin of massive O-type field stars: II. Field O stars as runaways.} \aap, \textbf{437}, 247 

\bibitem[Delgado et al.(2006)]{Delgado06} \textsc{Delgado, A.~J., 
Gonzalez-Martin, O., Alfaro, E.~J., \& Yun, J.~L.} 2006  {Multiwavelength analysis of the young open cluster NGC 2362.} \apj, submitted (astro-ph/0602487) 

\bibitem[Ezoe et al.(2006)]{Ezoe06} \textsc{Ezoe, Y., Kokubun, M., 
Makishima, K., Sekimoto, Y., \& Matsuzaki, K.} 2006  {Investigation of Diffuse Hard X-Ray Emission from the Massive Star-forming Region NGC 6334.} \apj, \textbf{638}, 860 

\bibitem[Feigelson et al.(2006)]{Feigelson06} \textsc{Feigelson, E., 
Townsley, L., Gudel, M., \& Stassun, K.} 2006  {X-ray Properties of Young Stars and Stellar Clusters.}  To appear in \textit{Protostars \& Planets V} (ed.\ B.\ Reipurth, D.\ Jewitt, \& K.\ Keil), Univ.\ Arizona Press (astro-ph/0602603)

\bibitem[Figuer{\^e}do et al.(2002)]{Figueredo02} \textsc{Figuer{\^e}do, 
E., Blum, R.~D., Damineli, A., \& Conti, P.~S.} 2002  {The Stellar Content of Obscured Galactic Giant H II Regions. IV. NGC 3576.} \aj, \textbf{124}, 2739 

\bibitem[Flaccomio et al.(2006)]{Flaccomio06} \textsc{Flaccomio, E., 
Micela, G., \& Sciortino, S.} 2006  {ACIS-I observations of NGC2264. Membership and X-ray properties of PMS stars.} \aap, accepted (astro-ph/0604243)

\bibitem[Franciosini et al.(2006)]{Franciosini06} \textsc{Franciosini, E., 
Pallavicini, R., \& Sanz-Forcada, J.} 2006  {XMM-Newton observations of the $\sigma$ Ori cluster. II. Spatial and spectral analysis of the full EPIC field.} \aap, \textbf{446}, 501 

\bibitem[Gagn{\'e} et al.(2005)]{Gagne05} \textsc{Gagn{\'e}, M., 
Oksala, M.~E., Cohen, D.~H., Tonnesen, S.~K., ud-Doula, A., Owocki, S.~P., 
Townsend, R.~H.~D., \& MacFarlane, J.~J.} 2005  {Chandra HETGS Multiphase Spectroscopy of the Young Magnetic O Star $\theta^1$ Orionis C.} \apj, \textbf{628}, 986 

\bibitem[Getman et al.(2006)]{Getman06} \textsc{Getman, K.~V., 
Feigelson, E.~D., Townsley, L., Broos, P., Garmire, G., \& Tsujimoto, M.} 
2006  {Chandra Study of the Cepheus B Star-forming Region: Stellar Populations and the Initial Mass Function.} \apjs, \textbf{163}, 306 

\bibitem[Hachisuka et al.(2006)]{Hachisuka06} \textsc{Hachisuka, K., et 
al.} 2006  {Water Maser Motions in W3(OH) and a Determination of Its Distance.} \apj, \textbf{645}, 337 

\bibitem[Hanson et al.(1997)]{Hanson97} \textsc{Hanson, M.~M., Howarth, 
I.~D., \& Conti, P.~S.} 1997  {The Young Massive Stellar Objects of M17.} \apj, \textbf{489}, 698 

\bibitem[Heyer \& Terebey(1998)]{Heyer98} \textsc{Heyer, M.~H., \& 
Terebey, S.} 1998  {The Anatomy of the Perseus Spiral Arm: 12CO and IRAS Imaging Observations of the W3-W4-W5 Cloud Complex.} \apj, \textbf{502}, 265 

\bibitem[Hofner et al.(2002)]{Hofner02} \textsc{Hofner, P., Delgado, H., 
Whitney, B., Churchwell, E., \& Linz, H.} 2002  {X-Ray Detection of the Ionizing Stars in Ultracompact H II Regions.} \apjl, \textbf{579}, L95 
 
\bibitem[Kraemer et al.(2003)]{Kraemer03} \textsc{Kraemer, K.~E., 
Shipman, R.~F., Price, S.~D., Mizuno, D.~R., Kuchar, T., \& Carey, S.~J.}
2003  {Observations of Star-Forming Regions with the Midcourse Space Experiment.} \aj, \textbf{126}, 1423 

\bibitem[Maercker et al.(2006)]{Maercker06} \textsc{Maercker, M., Burton, 
M.~G., \& Wright, C.~M.} 2006  {L-band (3.5 $\mu$m) IR-excess in massive star formation. II. RCW 57/NGC 3576.} \aap, \textbf{450}, 253 

\bibitem[Massey \& Hunter(1998)]{Massey98} \textsc{Massey, P., \& 
Hunter, D.~A.} 1998  {Star Formation in R136: A Cluster of O3 Stars Revealed by Hubble Space Telescope Spectroscopy.} \apj, \textbf{493}, 180 

\bibitem[Megeath et al.(2005)]{Megeath05} \textsc{Megeath, S.~T., Wilson, 
T.~L., \& Corbin, M.~R.} 2005  {Hubble Space Telescope NICMOS Imaging of W3 IRS 5: A Trapezium in the Making?} \apjl, \textbf{622}, L141 

\bibitem[Muno et al.(2006)]{Muno06} \textsc{Muno, M.~P., Law, C., 
Clark, J.~S., Dougherty, S.~M., de Grijs, R., Portegies Zwart, S., \& 
Yusef-Zadeh, F.} 2006  {Diffuse, Non-Thermal X-ray Emission from the Galactic Star Cluster Westerlund 1.} \apj, accepted (astro-ph/0606492) 

\bibitem[Nielbock et al.(2001)]{Nielbock01} \textsc{Nielbock, M., Chini, 
R., J{\"u}tte, M., \& Manthey, E.} 2001  {High mass Class I sources in M 17.} \aap, \textbf{377}, 273 

\bibitem[Oey et al.(2005)]{Oey05} \textsc{Oey, M.~S., Watson, A.~M., 
Kern, K., \& Walth, G.~L.} 2005  {Hierarchical Triggering of Star Formation by Superbubbles in W3/W4.} \aj, \textbf{129}, 393 

\bibitem[Persi et al.(1994)]{Persi94} \textsc{Persi, P., Roth, M., 
Tapia, M., Ferrari-Toniolo, M., \& Marenzi, A.~R.} 1994  {The young stellar population associated with the HII region NGC 3576.} \aap, \textbf{282}, 474 

\bibitem[Rathborne et al.(2002)]{Rathborne02} \textsc{Rathborne, J.~M., 
Burton, M.~G., Brooks, K.~J., Cohen, M., Ashley, M.~C.~B., \& Storey, 
J.~W.~V.} 2002  {Photodissociation regions and star formation in the Carina nebula.} \mnras, \textbf{331}, 85 


\bibitem[Rebull et al.(2006)]{Rebull06} \textsc{Rebull, L.~M., Stauffer, 
J.~R., Ramirez, S.~V., Flaccomio, E., Sciortino, S., Micela, G., Strom, 
S.~E., \& Wolff, S.~C.} 2006  {Chandra X-Ray Observations of Young Clusters. III. NGC 2264 and the Orion Flanking Fields.} \aj, \textbf{131}, 2934 

\bibitem[Sana et al.(2006)]{Sana06} \textsc{Sana, H., Gosset, E., 
Rauw, G., Sung, H., \& Vreux, J.~-.} 2006  {An XMM-Newton view of the young open cluster NGC 6231 -- I. The catalogue.} \aap, accepted (astro-ph/0603783)

\bibitem[Sanchawala et al.(2006)]{Sanchawala06} \textsc{Sanchawala, K., 
Chen, W.-P., Lee, H.-T., Nakajima, Y., Tamura, M., Baba, D., Sato, S., \& 
Chu, Y.-H.} 2006 {X-ray Emitting Young Stars in the Carina Nebula.} astro-ph/0603043 

\bibitem[Skinner et al.(2006)]{Skinner06} \textsc{Skinner, S.~L., 
Simmons, A.~E., Zhekov, S.~A., Teodoro, M., Damineli, A., \& Palla, F.} 
2006  {A Rich Population of X-Ray-emitting Wolf-Rayet Stars in the Galactic Starburst Cluster Westerlund 1.} \apjl, \textbf{639}, L35 

\bibitem[C.\ Smith et al.(2000)]{CSmith00} \textsc{Smith, C., Leiton, R., \& 
Pizarro, S.} 2000  {The UM/CTIO Magellanic Cloud Emission Line Survey (MCELS).} In \textit{Stars, Gas and Dust in Galaxies: 
Exploring the Links} (ed.\ D. Alloin, K.\ Olsen, \& G.\ Galaz), ASP Conf.~Ser.\ \textbf{221}, 83 

\bibitem[N.\ Smith et al.(2000)]{Smith00} \textsc{Smith, N., Egan, M.~P., 
Carey, S., Price, S.~D., Morse, J.~A., \& Price, P.~A.} 2000  {Large-Scale Structure of the Carina Nebula.} \apjl, \textbf{532}, L145 
 
\bibitem[N.\ Smith(2006a)]{Smith06a} \textsc{Smith, N.} 2006a  {A census of the Carina Nebula - I. Cumulative energy input from massive stars.} \mnras, \textbf{367}, 763 

\bibitem[N.\ Smith(2006b)]{Smith06b} \textsc{Smith, N.} 2006b  {The Structure of the Homunculus. I. Shape and Latitude Dependence from H2 and [Fe II] Velocity Maps of $\eta$ Carinae.} \apj, \textbf{644}, 1151 

\bibitem[R.\ Smith et al.(2001)]{Smith01} \textsc{Smith, R.~K., Brickhouse, 
N.~S., Liedahl, D.~A., \& Raymond, J.~C.} 2001  {Collisional Plasma Models with APEC/APED: Emission-Line Diagnostics of Hydrogen-like and Helium-like Ions.} \apjl, \textbf{556}, L91

\bibitem[Stassun et al.(2006)]{Stassun06} \textsc{Stassun, K.~G., van den 
Berg, M., Feigelson, E., \& Flaccomio, E.} 2006  {A Simultaneous Optical and X-ray Variability Study of the Orion Nebula Cluster. I. Incidence of Time-Correlated X-ray/Optical Variations.} \apj, accepted (astro-ph/0606079)

\bibitem[Tieftrunk et al.(1997)]{Tieftrunk97} \textsc{Tieftrunk, A.~R., 
Gaume, R.~A., Claussen, M.~J., Wilson, T.~L., \& Johnston, K.~J.} 1997  {The H II/molecular cloud complex W3 revisited: imaging the radio continuum sources using multi-configuration, multi-frequency observations with the VLA.}
\aap, \textbf{318}, 931 

\bibitem[Townsley et al.(2003)]{Townsley03} \textsc{Townsley, L.~K., 
Feigelson, E.~D., Montmerle, T., Broos, P.~S., Chu, Y., \& Garmire, G.~P.} 
2003  {10 MK Gas in M17 and the Rosette Nebula: X-Ray Flows in Galactic H II Regions.}  \apj, \textbf{593}, 874
     
\bibitem[Townsley et al.(2004)]{Townsley04} \textsc{Townsley, L., 
Feigelson, E., Montmerle, T., Broos, P., Chu, Y.~-., Garmire, G., \& 
Getman, K.} 2004  {Parsec-Scale X-ray Flows in High-Mass Star-Forming Regions.} Proc.\ of \textit{X-ray and Radio Connections} Workshop,
Santa Fe, NM, February 2004, astro-ph/0406349 

\bibitem[Townsley et al.(2005)]{TownsleyIAUS227} \textsc{Townsley, L.~K., 
Broos, P.~S., Feigelson, E.~D., \& Garmire, G.~P.} 2005  {Parsec-scale X-ray flows in high-mass star-forming regions.}  In \textit{IAU Symp.~227,
Massive Star Birth -- A Crossroads of Astrophysics} (ed.\ R.~Cesaroni, E.~Churchwell, M.~Felli, \& C.~M.~Walmsley), p.\ 297.  Cambridge University Press (astro-ph/0506418)

\bibitem[Townsley et al.(2006a)]{Townsley06a} \textsc{Townsley, L.~K., Broos,
P.~S., Feigelson, E.~D., Brandl, B.~R., Chu, Y.-H., Garmire, G.~P., \&
Pavlov, G.~G.} 2006a  {A {\em Chandra} ACIS Study of 30 Doradus. I. Superbubbles and Supernova Remnants.} \aj, \textbf{131}, 2140

\bibitem[Townsley et al.(2006b)]{Townsley06b} \textsc{Townsley, L.~K., Broos, P.~S.,
Feigelson, E.~D., Garmire, G.~P., \& Getman, K.~V.} 2006b  {A {\em Chandra} ACIS Study of 30 Doradus. II. X-Ray Point Sources in the Massive Star Cluster R136 and Beyond.}  \aj, \textbf{131}, 2164

\bibitem[Tsujimoto et al.(2005)]{Tsujimoto05} \textsc{Tsujimoto, M., 
Townsley, L., Feigelson, E.~D., Broos, P., Getman, K.~V., \& Garmire, G.}
2005  {Chandra Observations of Galactic HII Regions; NGC~7538 \& W49A.}  \textit{Protostars and Planets V}, 8307 

\bibitem[Tsujimoto et al.(2006)]{Tsujimoto06} \textsc{Tsujimoto, M., Hosokawa, T., Feigelson, E.~D., Getman, K.~V., \& Broos, P.} 2006  {Hard X-rays from Ultracompact HII Regions in W49A.}  \apj, accepted

\bibitem[V\'azquez et al.(1996)]{Vazquez96} \textsc{V\'azquez, R.~A., Baume, 
G., Feinstein, A., \& Prado, P.} 1996  {Investigation on the region of the open cluster TR 14.} \aaps, \textbf{116}, 75 

\bibitem[Walborn et al.(2002a)]{Walborn02a} \textsc{Walborn, N.~R., et al.} 
2002a  {A New Spectral Classification System for the Earliest O Stars: Definition of Type O2.} \aj, \textbf{123}, 2754 

\bibitem[Walborn et al.(2002b)]{Walborn02b} \textsc{Walborn, N.~R., Danks, 
A.~C., Vieira, G., \& Landsman, W.~B.} 2002b  {Space Telescope Imaging Spectrograph Observations of High-Velocity Interstellar Absorption-Line Profiles in the Carina Nebula.} \apjs, \textbf{140}, 407 

\bibitem[J.\ Wang et al.(2006)]{Wang06} \textsc{Wang, J., Townsley, L.~K., Feigelson, E.~D., Getman, K.~V., Broos, P.~S., Garmire, G.~P., Tsujimoto, M.} 2006  {A Chandra Study of the NGC 6357 Star-forming Complex.} \apjs, submitted

\bibitem[Q.\ Wang et al.(2006)]{DWang06} \textsc{Wang, Q.~D., Dong, H., \& 
Lang, C.} 2006  {The Interplay between Star Formation and the Nuclear Environment of our Galaxy: Deep X-ray Observations of the Galactic Center Arches and Quintuplet Clusters.} \mnras, accepted (astro-ph/0606282)

\bibitem[Wolk et al.(2006)]{Wolk06} \textsc{Wolk, S.~J., Spitzbart, 
B.~D., Bourke, T.~L., \& Alves, J.} 2006  {X-ray and IR Point Source Identification and Characteristics in the Embedded, Massive Star-Forming Region RCW 38.} \apj, accepted (astro-ph/0605096)



\end{thebibliography}
\end{document}